\documentclass[iop]{emulateapj}

\usepackage{natbib,epsfig,siunitx,url,graphicx,amsmath,footnote,float,xcolor,cases}
\usepackage[colorlinks=true,linkcolor=blue,citecolor=blue]{hyperref}

\begin{document}

\submitted{The Planetary Science Journal; accepted}

\title{Embryo formation with GPU acceleration: reevaluating the initial conditions for terrestrial accretion}

\author{Matthew S. Clement\altaffilmark{1,2}, Nathan A. Kaib\altaffilmark{2}, \& John E. Chambers\altaffilmark{1}}

\altaffiltext{1}{Earth and Planets Laboratory, Carnegie Institution
for Science, 5241 Broad Branch Road, NW, Washington, DC
20015, USA}
\altaffiltext{2}{HL Dodge Department of Physics Astronomy, University of Oklahoma, Norman, OK 73019, USA}
\altaffiltext{*}{corresponding author email: mclement@carnegiescience.edu}

\begin{abstract}

The solar system's terrestrial planets are thought to have accreted over millions of years out of a sea of smaller embryos and planetesimals.  Because it is impossible to know the surface density profile for solids and size frequency distribution in the primordial solar nebula, distinguishing between the various proposed evolutionary schemes has been historically difficult.  Nearly all previous simulations of terrestrial planet formation assume that Moon to Mars massed embryos formed throughout the inner solar system during the primordial gas-disk phase. However, validating this assumption through models of embryo accretion is computationally challenging because of the large number of bodies required. Here, we reevaluate this problem with GPU-accelerated, direct N-body simulations of embryo growth starting from $r\sim$100 km planetesimals.  We find that embryos emerging from the primordial gas phase at a given radial distance already have masses similar to the largest objects at the same semi-major axis in the modern solar system.  Thus, Earth and Venus attain $\sim$50$\%$ of their modern mass, Mars-massed embryos form in the Mars region, and Ceres-massed objects are prevalent throughout asteroid belt.  Consistent with other recent work, our new initial conditions for terrestrial accretion models produce markedly improved solar system analogs when evolved through the giant impact phase of planet formation.  However, we still conclude that an additional dynamical mechanism such as giant planet migration is required to prevent Earth-massed Mars analogs from growing.
\end{abstract}

\section{Introduction}

 Since advances in computing power led to the widespread availability of open source symplectic integrators \citep{wisdomholman,duncan98,chambers99}, numerous theoretical studies have been dedicated to understanding the origins of the solar system's terrestrial architecture \citep[for recent reviews on the topic see:][]{izidoro18_book_review,ray18_rev}.  Broadly speaking, a successful model for the inner solar system must reconcile the differences in mass distribution and orbital excitation between the modern system and the presumed primordial solar nebula \citep[e.g.:][]{mmsn}.  Of particular interest in the literature are the order of magnitude differences between the masses of Mercury and Mars and the neighboring Earth and Venus \citep{chambers01,ray09a,lykawka19}, the four orders of magnitude mass disparity between the asteroid belt and planetary regimes \citep{obrien07,izidoro15,clement18_ab}, and the plethora of high orbital eccentricities and inclinations in the asteroid belt \citep{petit01,morby10,deienno16}. While there are many compelling models and solutions to these issues, non-uniform disk conditions \citep{hansen09,iz14}, the influence of the giant planets \citep{walsh09,ray09a,lykawaka13,bromley17} and different modes of accretion \citep{levison15,morby15,draz16} feature prominently in most proposed evolutionary scenarios.  However, many models are based on an assumption that Moon to Mars massed planet-forming embryos were abundant throughout the terrestrial disk \citep{koko_ida02} that is in conflict with modern, high-resolution simulations of embryo formation \citep{carter15,walsh19}.  This is partially a consequence of the fact that, due to limits in computing power, the various phases of planet growth (planetesimal formation, embryo formation and the giant impact phase) are often treated separately.  We briefly summarize each phase of accretion below \citep[not discussed here, but still relevant for the solar system are gas accretion and planet migration; see reviews in:][]{morb12,ray18_rev}:
 
 \subsection{Planetesimal Formation}
 
Large infrared excesses in observed proto-planetary nebulae \citep{brice01} imply that the majority of the solid mass in young planet forming disks is concentrated in sub-micron sized dust grains.  Population studies of gaseous disk ages \citep{haisch01} indicate that they do not persist for longer than a few Myr.  Because ages of iron meteorites indicate that their primitive parent bodies accreted rapidly \citep[$\lesssim$5 Myr, apparently pre-dating the appearance of chondrules:][]{kleine05}, the transition from dust to 10-100 kilometer-scale planetesimals must have occurred rapidly, while gas was still present in the system.  Additionally, surveys of proto-$stellar$ disks indicate that they possess significantly higher dust masses than proto-$planetary$ disks \citep{tobin20}.  This result seems to imply that the conversion of dust to larger solid bodies occurs rapidly, and in conjunction with the earliest epochs of star formation.  

Dust grains can grow via various processes including coagulation \citep{xiang19}, aggregation and compaction \citep{wetherill80_rev,dominik07}.  However, explaining growth beyond meter-scales \citep[the so-called meter-barrier:][]{weidenschilling77a,birnstiel12} is difficult because millimeter sized bodies experience significant aero-dynamic drag, causing them to orbit at sub-Keplerian velocities and migrate inward \citep{whipple72}.  One intriguing solution to this issue might be direct formation via gravitational collapse.  If dust particles are sufficiently concentrated relative to the gas, they can clump together and form $D\sim$100km planetesimals rapidly via gravitational instability \citep{johansen15}.  While the ``streaming instability'' scenario offers a compelling resolution to the meter-barrier problem, the specific properties (radial location, final masses and formation time) of the resulting primordial generation of planetesimals are dependent on unconstrained disk parameters \citep{draz16,carrera17,abod19}.

 \subsection{Embryo Formation}
 
 Planetesimals continue to grow larger throughout the gas disk phase by direct accretion of other planetesimals \citep{wetherill93,koko_ida_98} and inward-drifting $\sim$meter-scale ``pebbles'' \citep[the radial flux of which is still debated:][]{johansen10,chambers16,ida16}.  This growth phase is highly efficient as long as the relative velocities in the region are low.  The largest local planetesimal gravitationally focuses \citep{chambers06} the incoming flux of planetesimals and pebbles, causing runaway growth to ensue \citep{koko_ida_96}.  Eventually, the oligarch planetesimals grow large enough to excite the orbits of nearby planetesimals and inhibit accretion \citep{koko_ida00}.  Growth can continue beyond this stage if the pebble flux is great enough \citep{lambrechts14b}, as pebbles are sufficiently small for gas disk interactions to damp their orbits.  Eventually, pebble accretion shuts off when an embryo grows large enough \citep[``pebble isolation mass:''][]{morb_nesvory_12} to induce a pressure gradient in the disk that prohibits inward pebble drift.  Recent high-resolution simulations in \citet{wallace19} uncovered an additional growth mode that occurs in the vicinity of massive oligarchs as small planetesimals are stacked inside of nearby first order mean motion resonances (MMR); thereby accelerating their growth towards intermediate masses and inducing a corresponding bump in the planetesimal size frequency distribution (SFD).
 
 \subsection{Giant Impact Phase}
 
 In the final phase of terrestrial growth, the embryo and planetesimal populations emerging from the gas disk collisionally assemble over $\sim$100 Myr timescales \citep{wetherill78,chambers01,ray09a}.  By and large, numerical models of the ``giant impact'' phase report timescales for the epoch's completion consistent with the geologically inferred timing of the Moon-forming impact \citep[$\sim$ 50-150 Myr;][]{kleine09}.  Moreover, it is within this ultimate stage of accretion that many authors have proposed solutions to the aforementioned mass and orbital excitation disparities in the inner solar system.  Accounting for hit-and-run collisions \citep{chambers13,clement18_frag} and dynamical friction induced by small bodies \citep{obrien06,ray06} can result in analog terrestrial planets with more realistic, dynamically cold orbits.  Furthermore, the giant planet instability \citep[the ``Nice Model'' of][]{Tsi05,levison08,nesvorny12} is typically invoked to explain the asteroid belt's excited state \citep{roig15,deienno16,deienno18} and (at least some of) its primordial depletion \citep{morby10,clement18_ab}.  Reconciling the Earth-Mars mass disparity, however, has led to the development of a multitude of different models.  It should also be mentioned here that collisional fragmentation \citep{chambers13} plays a role in the giant impact phase of terrestrial growth, the degree to which is a topic of continued debate \citep[for recent works espousing differing viewpoints, we direct the reader to:][]{clement18_frag,deienno19,kobayashi19}.  Specifically, a large, fragmenting collision \citep{asphaug14} is a potentially viable explanation for Mercury's large, iron-rich core \citep{jackson18}.
 
 \subsection{Small Mars Problem}
 
 Solutions to the small Mars problem \citep{wetherill91} generally fall in to one of two categories.  In the first class of models \citep[e.g.:][]{hansen09,ray17sci}, the outer terrestrial disk is already depleted during the primordial gas-disk phase, and the terrestrial planets form out of a narrow annulus of material.  In one such scenario \citep[the so-called ``Grand Tack'' model of][]{walsh11,jacobson14,walsh16}, the inner disk is truncated when Jupiter migrates into, and subsequently back out of the terrestrial region.  Thus, Mars forms rapidly as a ``stranded embryo'' \citep{Dauphas11,iz14}. In contrast, the second class of models invoke a dynamical mechanism to starve the region of material during the planet formation process \citep{ray09a}.  Typically, the influence of Jupiter and Saturn perturbs objects in the region, and inhibits Mars' formation.  Resonance sweeping or crossing \citep{lykawaka13,bromley17}, primordially excited giant planet orbits \citep{ray09a,lykawka19} and the Nice Model instability \citep{clement18} have all been shown to substantially restrict Mars' growth.  In particular, the ``Early Instability'' scenario argues that a Nice Model timed $\sim$1-5 Myr \citep{clement18_frag} after gas disk dispersal explains Mars' rapid inferred geologic formation time \citep{Dauphas11,kruijer17_mars}.  However, each model should be considered in the appropriate context given the fact that Mercury's low mass and orbit are still very low probability outcomes of numerical simulations \citep{ls14,clement19_merc}.
 
 \subsection{This Work}
  
With few exceptions, the aforementioned N-body studies of the giant impact phase all place large embryos throughout the terrestrial disk and modern asteroid belt.  However, such initial conditions are at odds with semi-analytic predictions of oligarchic growth \citep{koko_ida_98,koko_ida00}, as well as recent high-resolution studies of embryo formation within gaseous disks \citep{carter15,wallace19,walsh19}.  In particular, it appears unlikely that the primordial asteroid belt region attained such an advanced evolutionary state during the gas disk phase.  In this paper we follow the complete growth of the terrestrial system starting from $r\sim$100 km planetesimals accreting in a decaying gas disk \citep{morishma10}.  It should be noted that, as a tangentially related alternative to our proposed scenario, self-consistent pebble accretion (not considered in our work) simulations \citep[e.g.:][]{morby15} form embryos directly throughout the terrestrial disk.

Our current study is perhaps most similar to the recent work of \citet{carter15} and \citet{walsh19}, and we compare our results with both authors' findings throughout this manuscript.  While \citet{carter15} used a parallelized N-body code \citep[$PKDGRAV$:][]{pkdgrav1,pkdgrav2} with inflated planet radii and \citet{walsh19} employed a Lagrangian integrator and tracer particles \citep[the $LIPAD$ code:][]{lipad}, we opt for a direct, GPU accelerated N-body scheme \citep[$GENGA$:][]{genga} that fully resolves close encounters.  Notably, we investigate the effects of Jupiter and Saturn's presence, and use  self-interacting planetesimals for much of the runaway growth phase.  While we leave the full-resolution evolution of our generated distributions of embryos and planetesimals through the giant impact phase to a future paper, we perform an additional suite of simplified simulations of the final stage of accretion for a first order approximation of the final system architectures.

\section{Methods}

To achieve sufficiently high particle resolution throughout the terrestrial disk, we begin by numerically integrating different radial annuli separately.  As oligarchic growth ensues in each annulus, we begin to merge our simulations; combining and interpolating between individual annuli until the entire terrestrial disk is assembled in a single simulation at the $t=$1 Myr.  For each of the simulations that include gas drag, we use the GPU parallelized hybrid symplectic integrator $GENGA$ \citep{genga}.  $GENGA$ is based on the $Mercury$ code of \citet{chambers99}, runs on all $Nvidia$ GPUs, and is available to the public in an open source format.  In all of our simulations we employ a time-step equal to $\sim$5$\%$ that of the shortest orbital period, remove objects that pass within 0.1 au of the Sun \citep[common practice in N-body studies of planet formation, see][]{chambers01}, and consider particles ejected from the system at 100 au.  We also incorporate a simple, analytic gas disk model \citep{morishma10} that mimics the effects of aerodynamic drag \citep{adachi76}, $e/i$ damping induced by tidal interactions between the gas disk and proto-planets \citep{tanaka02}, and the global nebular gravitational \citep{nagasawa00} force by applying an additional acceleration to each body after the Keplerian drift kicks.  In this model, gas dissipates exponentially in time and uniformly in space as:

\begin{equation}
	\Sigma_{gas}(r,t) = \Sigma_{gas}(1\medspace au,0)\bigg(\frac{r}{1\medspace au}\bigg)^{-\alpha}exp\bigg(-\frac{t}{\tau_{gas}}\bigg)
\end{equation}

While previous authors \citep{morishma10,walsh19} have investigated different gas densities and decay times, because we are constricted by the availability of GPUs, we limit our study to the nominal minimum mass solar nebula \citep[MMSN:][]{hayashi81}.  Thus, for all of our integrations, we use a decay time of $\tau_{gas}=$ 3 Myr, an initial surface density of $\Sigma_{gas}(1\medspace au,0)=$ 2,000 $g\thinspace cm^{-2}$, and set $\alpha=$1 \citep[based on the nominal $\alpha$ of][]{morishma10} .

\subsection{Runaway growth in the inner disk: 0-100 Kyr}
\label{sect:meth_ann}

We begin by following the evolution of five, 0.1$M_{\oplus}$ annuli located at 0.5, 1.0, 1.5, 2.0 and 3.0 au (see table \ref{table:ann}) for 100 Kyr.  Each annulus is composed of 5000, fully self-gravitating, equal-massed planetesimals on nearly circular, co-planar orbits \citep[we draw initial eccentricities and inclinations from Rayleigh distributions with $\sigma_{e}=$0.002 and $\sigma_{i}=$0.2$^{\circ}$ as described in][]{ida90, koko_ida_98}.  Thus, assuming a nominal planetesimal density of 3.0 $g\thinspace cm^{-3}$, each object has D$\approx$200 km.  Our annuli are derived from a terrestrial disk (0.5-4.0 au) that contains 5 $M_{\oplus}$ of solid material with a surface density profile that falls off radially as $r^{-3/2}$ \citep[consistent with studies of terrestrial planet formation:][]{wetherill96,chambers01,ray09a,walsh19}.  To account for edge effects, we employ a boundary condition similar to \citet{koko_ida00}.  If a particle's semi-major axis passes beyond 0.005 au of either annulus edge, it is removed from the simulation and a new object is added at the opposite edge with inclination and eccentricity drawn from the annulus' in situ distributions.  The largest error introduced by this choice of boundary condition is an artificial damping of orbits near the boundary edge because the objects with higher eccentricities are being removed. However, the gas-driven migration timescale for $r=$ 100 km planetesimals \citep[for example:][]{weidenschilling77a} is greater than the nebular lifetime for our gas disk model \citep{morishma10}.  Moreover, radial scattering due to close encounters is significantly subdued in this phase of our simulations as a result of eccentricities being highly damped.  Therefore, in practice, this exchange of particles rarely occurs because orbits are highly damped.

\begin{table}
\centering
\begin{tabular}{c c c c c}
\hline
Annulus & $a_{in}$ (au) & $a_{center}$ (au) & $a_{out}$ (au) & $N_{f}/N_{i}$ \\
\hline
1 & 0.48 & 0.50 & 0.52  & 0.06 \\	
2 & 0.9745 & 1.0 & 1.0255 & 0.27 \\	
3 & 1.4685 & 1.5 & 1.5315 & 0.51 \\	
4 & 1.9635 & 2.0 & 2.0365 & 0.68 \\
5 & 2.9505 & 3.0 & 3.0495 & 0.88\\
\hline	
\end{tabular}
\caption{Summary of annulus edges and centers for simulations of oligarchic growth from 0-100 Kyr.  Each annulus contains 5,000 fully self-gravitating objects and has a total mass of 0.1 $M_{\oplus}$.  Note that the different annulus widths are a result of the $\Sigma \propto r^{-3/2}$ surface density profile.}
\label{table:ann}
\end{table}

\subsection{Runaway growth in the outer disk: 100-1000 Kyr}
\label{sect:meth_interpolate}

\begin{figure*}
	\centering
	\includegraphics[width=.9\textwidth]{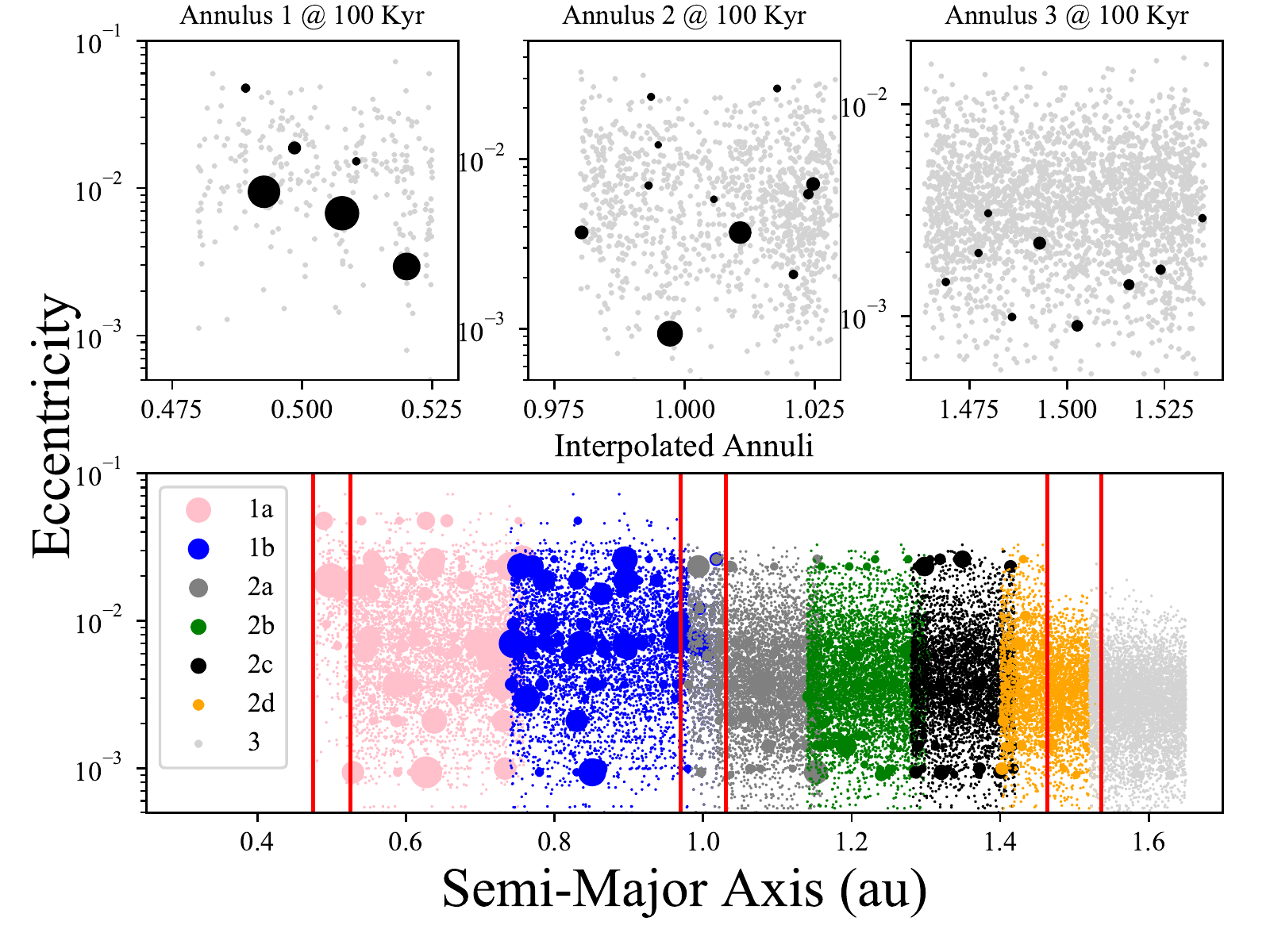}
	\caption{$a/e$ plot of our three innermost annuli (top panels) and 7 new interpolated annuli (bottom panel, table \ref{table:new}) at $t=$ 100 Kyr.  The edges of the three original initial annuli (1, 2 and 3) from which these new disk conditions are derived are denoted with vertical red lines.  The size of each point is scaled to the object's mass, and each color represents a different simulation annuli (by design, the annuli slightly overlap).  Note that, due to our method of interpolating between annuli (section \ref{sect:meth_interpolate}) by drawing from the in-situ distribution of orbital elements in the adjacent two annuli, the resulting interpolated particles possess discrete eccentricity values.}
		\label{fig:interp}
\end{figure*}

Since runaway growth ensues rapidly \citep{koko_ida_98} at small radial distances, the total number of particles in our $r=$ 0.5, 1.0 and 1.5 au annuli is small enough at $t=$ 100 Kyr ($N_{f}/N_{i}=$ 0.06, 0.27 and 0.51, respectively) to begin interpolating between annuli.  We generate 7 wider annuli (table \ref{table:new} and figure \ref{fig:interp}) based on the mass and orbital distributions within our inner 3 annuli after 100 Kyr of evolution. Combined, these new simulations span the entire radial range of 0.48-1.65 au.  To accomplish this interpolation, we first separate each annulus into three classes of particles: oligarch embryos (the most massive object in the annulus), proto-embryos (all other objects larger than 25 times the initial planetesimal mass), and planetesimals.  These class divisions are purely used for our interpolation methodology, and do not affect how the particles are treated in the actual integration.  In the intra-annulus regions, we assume that the percentage of total mass concentrated in embryos (both oligarchs and proto-embryos) and planetesimals is equal to the arithmetic mean of the corresponding percentages in the neighboring annuli.  We then place new oligarch embryos in the region with linearly decreasing masses, and semi major axes that maintain a surface density profile proportional to $r^{-3/2}$.  Inclinations and eccentricities for the new oligarch embryos are chosen at random from the original embryo distributions.   Finally, we add new proto-embryos and planetesimals by randomly drawing masses, eccentricities and inclinations from the respective distributions, and semi-major axes that maintain $\Sigma_{disk} \propto r^{-3/2}$.  Therefore, as radial distance increases in the intra-annulus regions, total embryo mass is concentrated in a greater number of smaller embryos.  All three classes of objects (planetesimals, proto-embryos and embryos) interact with one another gravitationally for this phase of our study.

To verify the effectiveness of our interpolation method, we generate a new, artificial, annulus 2 utilizing the outputs of annuli 1 and 3 at $t=$ 100 Kyr.  When we compare our new artificial annulus with the actual state of annulus 2 at 100 Kyr, we find that the two systems are remarkably similar.  Annulus 2 contains 1,378 total particles at this stage of evolution, and our interpolation method creates a system of 1,354 particles.  Additionally, our scheme overestimates the mass of the oligarch in annulus 2 by just 3.9$\%$.  The ratio of total embryo (oligarchs and proto-embryos) to planetesimal mass in the actual annulus 2 is 0.54, as compared to 0.60 in our artificial system.  The only manner in which the two systems are significantly dissimilar is in the mass distribution of the proto-embryos.  Because we are interpolating over such a wide radial range, our artificial system contains around twice as many proto-embryos that are, on average, half as massive as those in the real annulus 2.  As this issue is lessened when the intra-annulus distance is reduced, we argue that our interpolation prescription is an adequate means for accelerating our calculations while minimizing additional error terms.

\begin{table}
\centering
\begin{tabular}{c c c c}
\hline
Annulus & $a_{in}$ (au) & $a_{out}$ (au) & $N_{part}$ \\
\hline
1a & 0.48 & 0.76 & 4984 \\	
1b & 0.74 & 1.02 & 5106 \\	
2a & 0.98 & 1.16 & 5622 \\	
2b & 1.14 & 1.30 & 5123 \\
2c & 1.28 & 1.42 & 3852\\
2d & 1.40 & 1.54 & 3758 \\
3a & 1.52 & 1.65 & 5355\\
\hline	
\end{tabular}
\caption{Summary of annulus edges and total particle numbers for simulations of embryo growth from 0.1-1.0 Myr.}
\label{table:new}
\end{table}

Through a process of trial and error, we determine the largest radial bins $GENGA$ can efficiently integrate with fully-interacting particles.  These seven new, slightly overlapping annuli (table \ref{table:new} and figure \ref{fig:interp}) are integrated for 900 Kyr utilizing the same gas disk and boundary conditions described in section \ref{sect:meth_ann}.  Conversely, our two original outermost annuli (2.0 and 3.0 au) are integrated up to $t=$1 Myr as is.

\subsection{Gas dispersal and the influence of Jupiter and Saturn: 1-3 Myr}

\begin{figure}
	\centering
	\includegraphics[width=.5\textwidth]{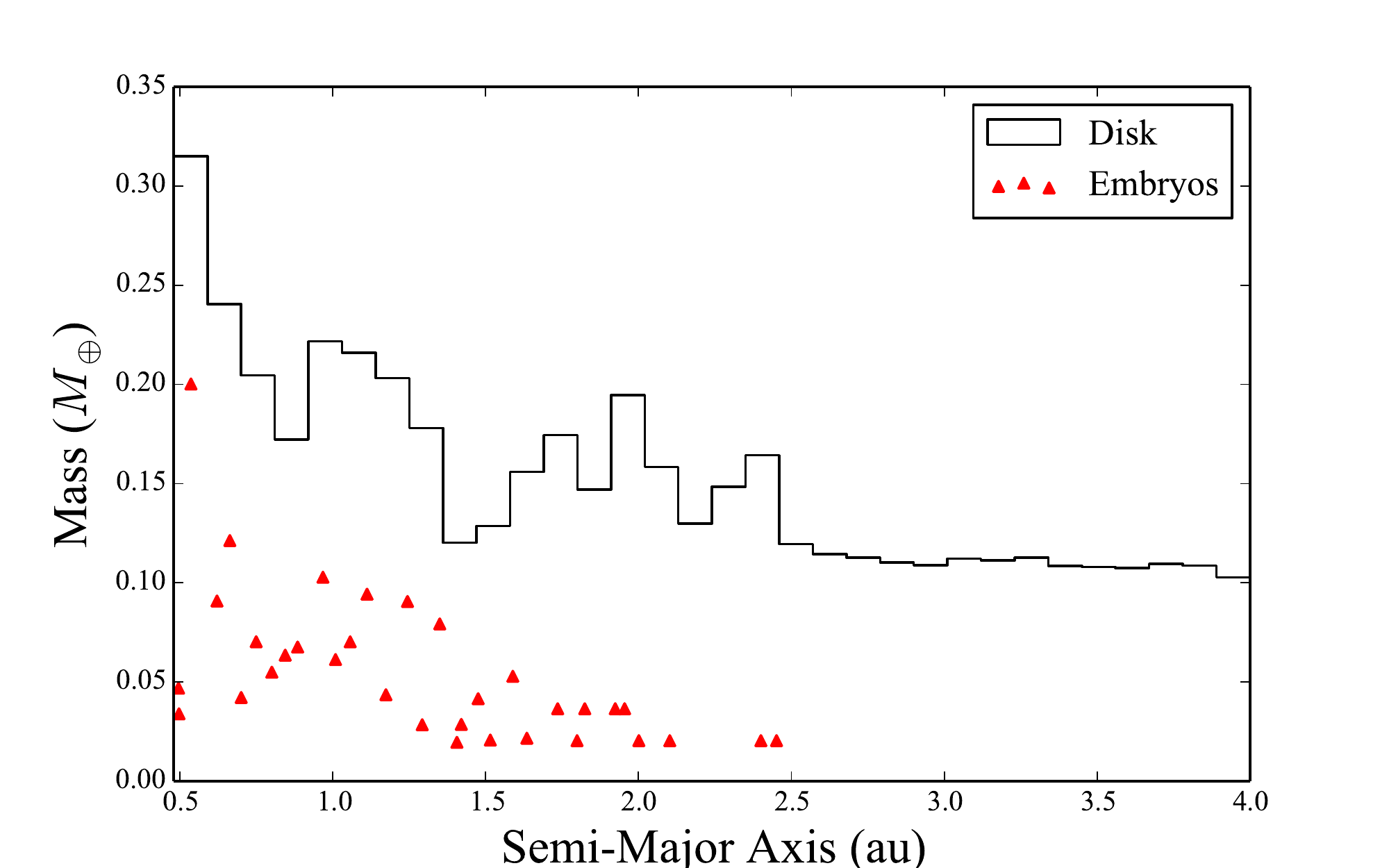}
	\caption{State of the terrestrial disk at 1 Myr; the point where we combine all annuli and add giant planets.  At this stage, the system contains 43,608 total disk particles.  The state of these systems at $t=$ 3 Myr is depicted in figure \ref{fig:ic_comp}  The black line represents the total disk mass profile, while the red triangles plot each individual embryo with $M>$0.01 $M_{\oplus}$.}
	\label{fig:1myr}
\end{figure}

At $t=$1 Myr, we combine all annuli in one large simulation containing 43,608 disk particles using the same interpolation method described in section \ref{sect:meth_interpolate} (however, all objects with $a>$2.0 au and $M>$25$M_{init}$ are treated as proto-embryos).  Figure \ref{fig:1myr} plots the radial mass profile of our fully constructed terrestrial formation disk. Because we do not include the 3.0$<r<$4.0 au region of the disk in any of our $t=$ 0-1 Myr integrations given the lengthy timescales for collisions to occur at large radial distances, we approximate this section of the disk in these simulations with planetesimals five times as massive as our initial planetesimals.  Since this region is quickly excited and eroded by perturbations from the giant planets following gas disk dispersal \citep[e.g.:][]{ray06}, we populate the region with unrealistically large asteroids only for the purposes of modeling the exterior mass' effect on the interior disk regions.  To further accelerate our calculation, we also  treat all planetesimals ($M<25M_{init}$) as semi-interacting (interact gravitationally with the embryos, but not with one another) for this phase of our study \citep[e.g.:][]{ray09a}.  However, all particles still feel the effects of the decaying gas disk.

At this stage of analysis, we begin to consider the effects of the growing gas giants with 3 separate models:

\begin{itemize}
	\item No giant planets (NOJS)
	\item Jupiter and Saturn each with $M=$ 8.0 $M_{\oplus}$ (8JS).
	\item Jupiter and Saturn begin with $M=$ 1.0 $M_{\oplus}$ and grow logarithmically to 95$\%$ their modern masses at $t=$ 3 Myr (GROW).
\end{itemize}

In each case, the giant planets are placed on near-circular orbits in mutual 3:2 \citep[$a_{Jup}=$5.6 au, consistent with the planets' presumed per-instability orbits, see][]{nesvorny11,deienno17} MMR \citep[e.g.:][]{lee02,clement18}.  We then integrate each system up to $t=$ 3 Myr as described above.  While none of these giant planet mass configurations are akin to that of the actual solar system, we include them for the purposes of testing, to first order, Jupiter and Saturn's effect on this phase of terrestrial evolution.

\subsection{The giant impact phase: 3-200 Myr}
\label{sect:nbody}

While we plan to continue the full resolution study of our complete terrestrial disk (figure \ref{fig:1myr}) in a future paper, we present a suite of simplified CPU simulations of the giant impact phase here to briefly comment on the implications of our generated disks.  These integrations make use of the $Mercury6$ hybrid integrator \citep{chambers99}, employ a 6 day time-step, and remove bodies with $r>$ 100 au and $r<$ 0.1 au.  Modern massed versions of Jupiter and Saturn are placed in a 3:2 MMR as described above.  All embryos (here $M\geq$ 0.01 $M_{\oplus}$) from our $GENGA$ simulations are included as fully interacting bodies.  The remaining disk mass is replaced by 1,000, equal-mass, semi-interacting planetesimals with orbits drawn randomly from the remaining $GENGA$ particles such that $\Sigma_{disk} \propto r^{-3/2}$ is maintained.  We perform 24 (as these are run on a cluster with 24 CPU cores per node) separate simulations in this manner for our NOJS, 8JS and GROW disks.  Additionally, we completely remove all gas disk interactions in these simplified integrations.  While a step change in the masses of Jupiter and Saturn and gas abundance at $t=$3 Myr is obviously not realistic, we present these simulations here to provide a zeroth order approximation of the final system architectures.

It should be noted here that the remnant planetesimal population can significantly affect the system's evolution within the giant impact phase \citep[see][where the ratio of total embryo to planetesimal mass is varied]{ray06,ray07,jacobson14}.  Therefore, our study is inherently biased by our initial planetesimal masses.  If the first generation of planetesimals indeed formed large and rapidly \citep[e.g.:][]{morby09_ast,johansen15,dermott18}, then our $D=$ 200 km initial bodies might be realistic.  Therefore, we conclude our study with a discussion of how the remnant planetesimal SFD can affect growing embryos with an additional suite of 50 simplified simulations of terrestrial accretion.  These simulations are performed with the $Mercury6$ hybrid integrator as described above utilizing embryo and planetesimal distributions akin to those supposed in classic N-body studies of terrestrial planet formation \citep{chambers01,chamb_weth01,ray09a}.  Each simulation assumes a 5$M_{\oplus}$ disk with half its mass concentrated in 50 equal-mass embryos, and the other 50$\%$ distributed equally between either 1,000 or 2,000 planetesimals (25 integrations each).  Semi-major axes are selected to achieve $\Sigma_{disk} \propto r^{-3/2}$, while eccentricities and inclinations are drawn from Rayleigh distributions ($\sigma_{e}=$0.002 and $\sigma_{i}=$0.2$^{\circ}$).

\section{Results and Discussion}

We present the results of our GPU-accelerated simulations of embryo growth in the subsequent sections \ref{sect:runaway}-\ref{sect:ics}.  The following sections, \ref{sect:tp_form}-\ref{sect:control}, discuss the outcomes of our additional, CPU-only, simulations of the giant impact phase of terrestrial planet formation.

\subsection{Oligarchic Growth}
\label{sect:runaway}

\begin{figure}
	\centering
	\includegraphics[width=.5\textwidth]{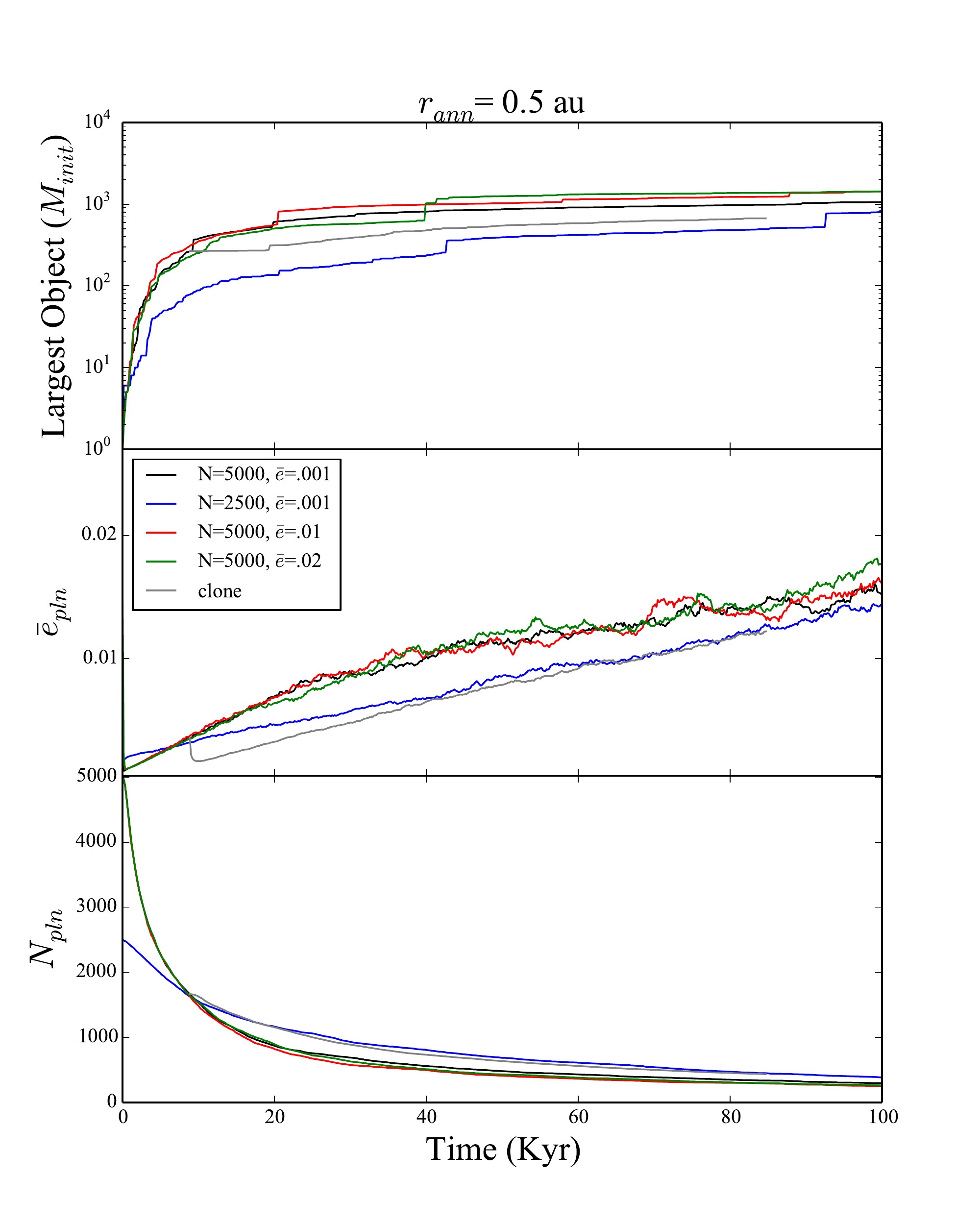}
	\caption{Comparison of varied initial conditions for our $r=$ 0.5 au annulus.  In the first four runs (color coded black, blue, red and green), the initial planetesimal eccentricities and total number of annulus particles are varied.  In the final simulation (``clone,'' grey line), the annulus is tripled in size once the total particle number drops by a factor of three.}
	\label{fig:emb_compare}
\end{figure}

We begin our analysis with a brief validation of our initial conditions and methods.  We perform four additional integrations of our innermost ($r=$ 0.5 au) annulus (plotted in figure \ref{fig:emb_compare}).  In two simulations, we increase the RMS eccentricity of our planetesimals ($\bar{e}_{pln}= $0.01 and 0.02; red and green lines in figure \ref{fig:emb_compare}, respectively) and verify that our results are independent of the particular initial orbits of the planetesimals.  Indeed, both simulations experience nearly identical runaway growth sequences.  This is because planetesimal orbits are rapidly damped to nearly zero eccentricity at the beginning of our simulations when the gas disk is particularly dense.  Next, we perform an integration where the annulus is represented by 2,500 equal-massed objects, rather than 5,000.  While the embryo growth sequence and planetesimal SFD in this run (blue line in figure \ref{fig:emb_compare}) are obviously different, we find that the net result at the end of the runaway growth phase is largely the same as in our nominal run in terms of $\bar{e}_{pln}$ and final embryo mass.  Finally, we scrutinize the effectiveness of our boundary condition by tripling the size of our nominal annulus once the total number of particles decreases by a factor of three.  This is accomplished by generating two, identical exterior annuli where the semi-major axis of each particle is shifted by the annulus width (0.04 au) while holding the other orbital elements constant.  The additional dynamical friction of the new surrounding planetesimals has the immediate effect of briefly damping planetesimal eccentricities (grey line in figure \ref{fig:emb_compare}), however the net result of the oligarchic growth scheme in terms of $M_{iso}$ and $T_{grow}$ is the same as in our nominal run after 100 Kyr.

\begin{figure}
	\centering
	\includegraphics[width=.5\textwidth]{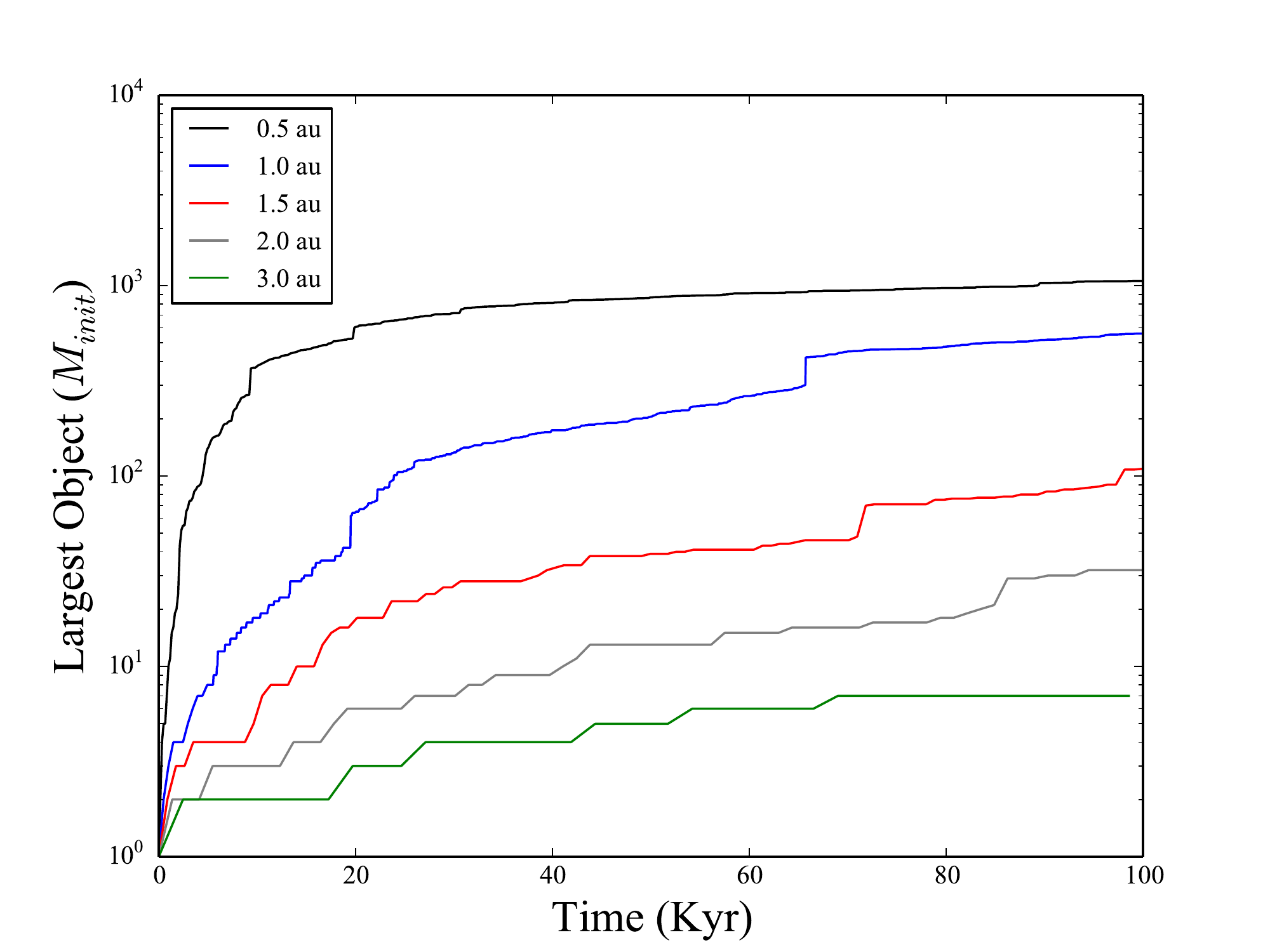}
	\caption{Growth of the largest object in each of our five initial annuli.  Oligarchic growth ensues rapidly in the inner disk, while our outermost annulus experiences few accretion events in the first 100 Kyr of evolution.}
	\label{fig:emb_time}
\end{figure}

The mass evolution from $t=$ 0-100 Kyr of the largest object in each of our five initial annuli (1-5, table \ref{table:ann}) is plotted in figure \ref{fig:emb_time}.  Runaway growth is self-limiting in the sense that it is effectively only dependent on the available mass to be accreted (a function of the local planetesimal surface density and, not considered here, the inward pebble flux) and the RMS eccentricity \citep[e.g.:][]{lissauer87} in the region as:

\begin{equation}
	\frac{1}{M}\frac{dM}{dt}\propto \Sigma_{pln} M^{1/3}\bar{e}_{pln}^{-2}
\end{equation}

The growth timescale while in the runaway regime is

\begin{equation}
	T_{grow} = \frac{M}{dM/dt}.
\end{equation}

Taking the ``kinetic gas'' approach \citep{wetherill80}, and assuming that dM/dt goes as

\begin{equation}
	\frac{dM}{dt} = \pi R^{2} \rho_{pln} \bar{v}_{pln} \bigg[1 + \bigg(\frac{v_{esc}}{\bar{v}_{pln}}\bigg)^{2}\bigg],
\end{equation}
(where $v_{esc}$ is the escape velocity at the surface of a growing embryo, $\rho_{pln}$ is the volume density of planetesimals and $\bar{v}_{pln}$ is their velocity dispersion) it can be shown that:

\begin{equation}
	T_{grow} \simeq 2 \times 10^4 \bar{e}_{pln}^{2}\bigg(\frac{\Sigma_{pln}}{\Sigma_{o}}\bigg)^{-1}\bigg(\frac{M}{10^{26}\medspace g}\bigg)^{1/3}\bigg(\frac{a}{1\medspace au}\bigg)^{2} yr.
	\label{eqn:tgrow}
\end{equation}
where $\Sigma_{o} =10\thinspace g\thinspace cm^{-2}$ at 1 au \citep[see][for a full derivation]{k_and_i95}.  Note that, here, the RMS planetesimal eccentricity is scaled by the reduced Hill radius:

\begin{equation}
	h_{r} = \bigg(\frac{M}{3M_{\odot}}\bigg)^{1/3}
	\label{eqn:miso}
\end{equation}

\begin{figure}
	\centering
	\includegraphics[width=.5\textwidth]{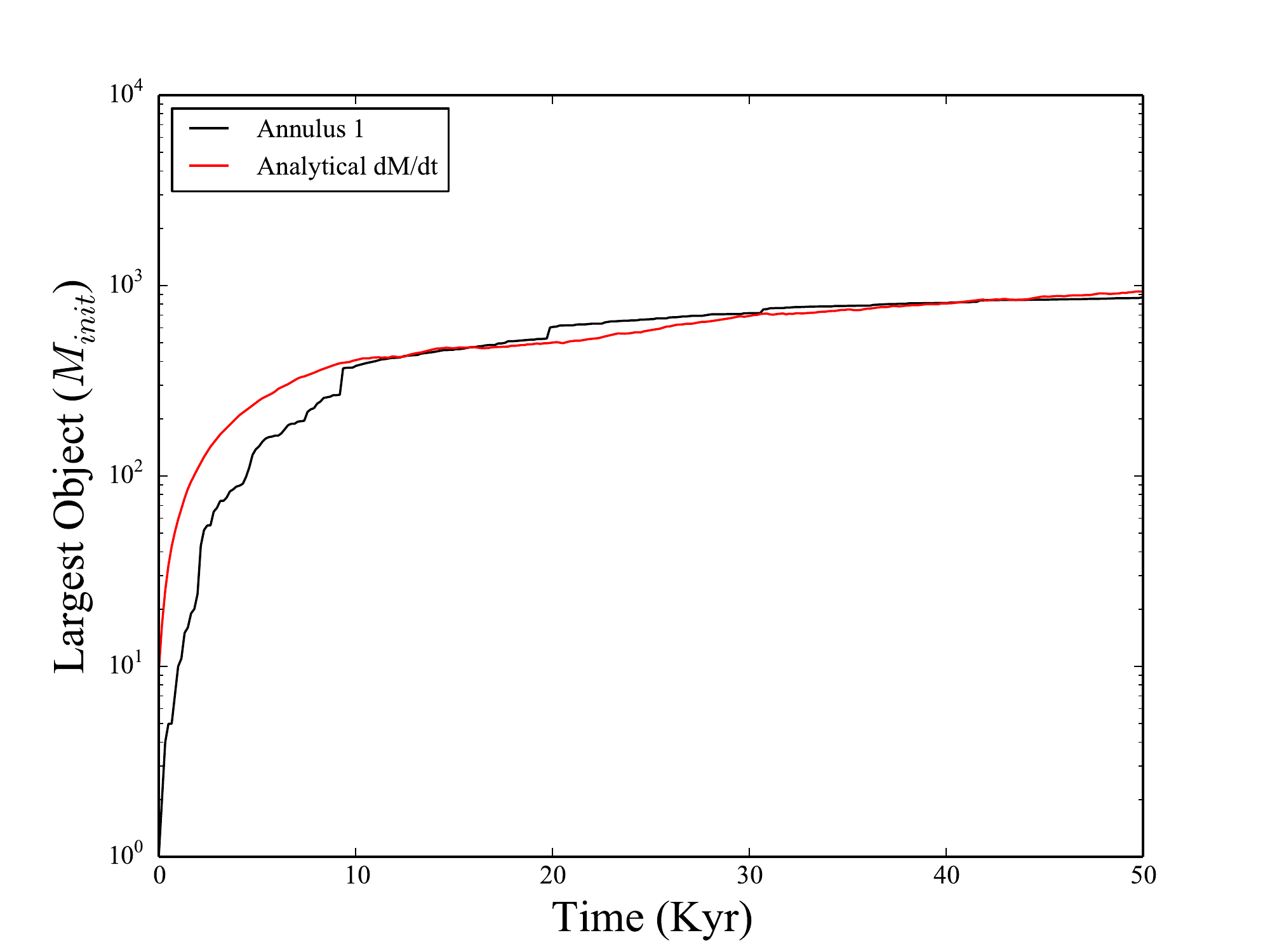}
	\caption{The same as figure \ref{fig:emb_time} for annulus 1, compared with analytical $dM/dt$ (equation \ref{eqn:tgrow}) utilizing the in-situ values of $\Sigma_{pln}$, $M$ and $\bar{e}$. As $T_{grow}<< \tau_{gas}$ in annulus 1, accretion is rapid and well characterized by the runaway growth regime.}
	\label{fig:added}
\end{figure}

\begin{figure}
	\centering
	\includegraphics[width=.5\textwidth]{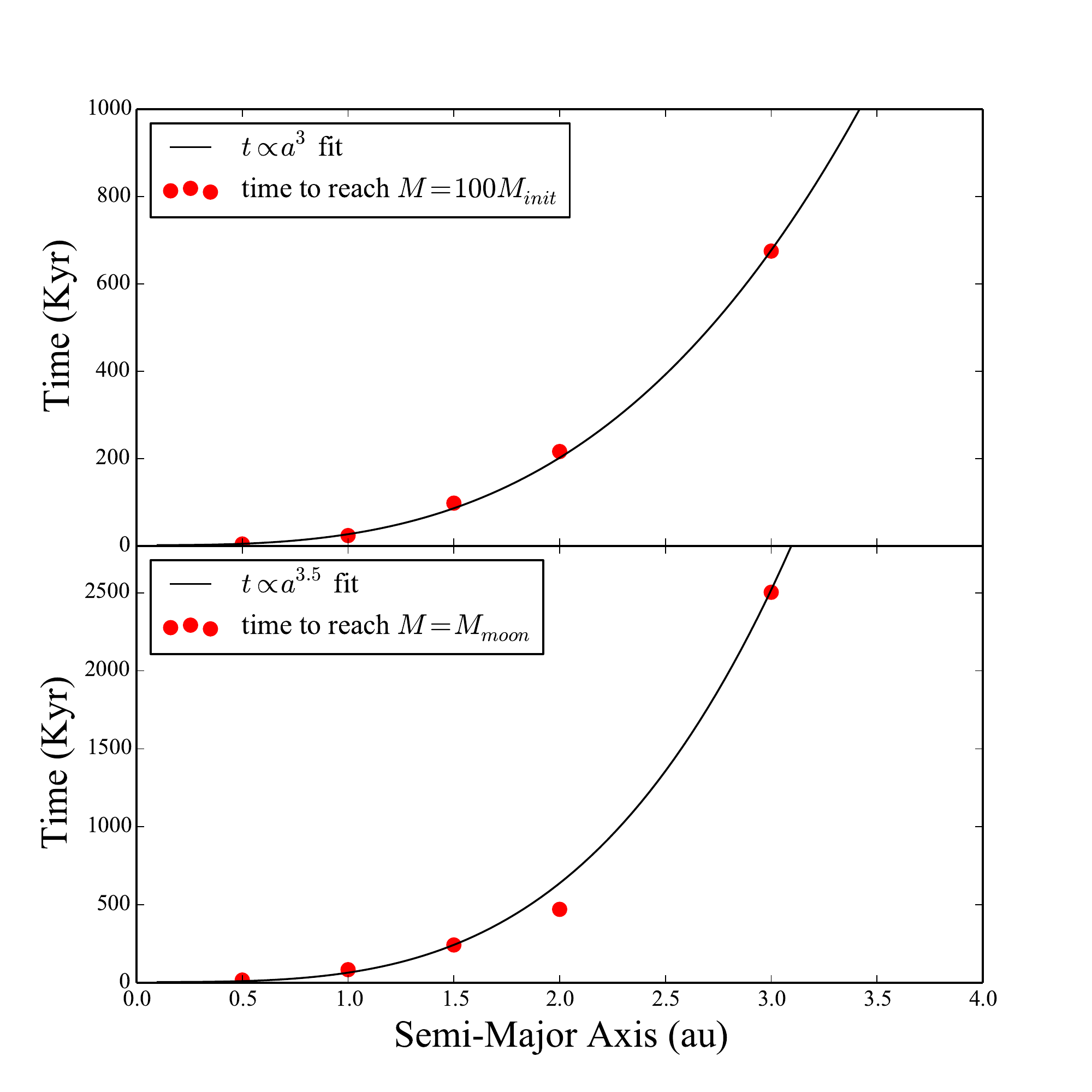}
	\caption{The time required for a planetesimal to increase in mass by two orders of magnitude (top panel) and the time to accrete a Moon massed embryo (bottom panel) for each of our 5 annuli.}
	\label{fig:powerlaw}
\end{figure}

The runaway growth regime is only relevant when $\Sigma_{pln}$ is large, and ample planetesimals are available to feed the growing embryo.  Therefore, this relationship (\ref{eqn:tgrow}) is typically cited in reference to growth towards the ``isolation mass'' \citep[e.g.:][]{koko_ida02,kobayashi13}:

\begin{equation}
	M_{iso} = 0.14 \chi^{3/2}\bigg(\frac{a}{1.5 au}\bigg)^{3/4}
\end{equation}
where $\chi$ is the scaling of the classic \citet{hayashi81} MMSN.  For $\chi=$1, this relationship predicts Mars massed embryos accreting at $\sim$1.5 au, and larger embryos in the asteroid belt region.  In practice, however, the isolation mass is likely never reached in the outer terrestrial disk as the timescales for giant planet growth and gas dispersal are significantly shorter than $T_{grow}$ for a$\gtrsim$1.5 au.  Thus, other dynamical processes likely begin perturbing this region long before $M_{iso}$ is reached.  In contrast to the outer disk, accretion in our innermost annuli is indicative of  runaway growth (equation \ref{eqn:tgrow} and figure \ref{fig:added}. For our purposes, however, we are most interested in the time required to accrete embryos of different masses as we seek to infer the conditions of the terrestrial disk around the time of nebular gas dispersal.  Figure \ref{fig:powerlaw} depicts this relationship for our five annuli for two different growth masses.  Assuming no evolution in $\bar{e}_{pln}^{2}$, from equation \ref{eqn:tgrow}, we would expect $T_{grow}$ in our $\Sigma \propto r^{-3/2}$ disk to scale as $a^{2}$ with increasing radial distance.  In our simulations, however, we find that it scales closer to $\sim a^{3}$.  Additionally, growth towards larger masses (bottom panel of figure \ref{fig:powerlaw}) is further curtailed in the outermost annulus and better fit by an $\sim a^{3.5}$ radial dependency.  These results are largely consistent with previous studies \citep{koko_ida02,chambers06,kobayashi13}, and analytical derivations incorporating the radial dependencies of $\bar{e}_{pln}$ (mainly a result of gas dynamics) and the isolation mass  \citep[e.g.:][infer an $a^{2.7}$ dependency]{koko_ida02}.  In a recent study similar to our current work, \citet{walsh19} report that $T_{grow}$ towards $M_{iso}$ scales as $\sim a^{3.6}$ in nominal MMSN models without collisional grinding.  Since $M_{iso}$, increases with radial distance (equation \ref{eqn:miso}), our measured times to reach a fixed embryo mass in different radial annuli are indicative of an even steeper scaling of $T_{grow}$ with $a$, particularly in our outermost annulus.  This is, at least partially, a result of our inclusion of the giant planets in two of our 1-3 Myr simulations.  While figure \ref{fig:powerlaw} plots the time to reach $M=M_{moon}$ for annulus 5 (table \ref{table:ann}) as the average of all 3 simulations, we note that this time $\sim$600 Kyr shorter in our NOJS run than in our other two runs.  Thus, perturbations from the growing giant planets, though significantly damped in the gas disk phase, are still sufficient to moderately excite orbits and limit accretion events in the asteroid belt region.

\subsection{Initial Conditions after Nebular Gas Dissipation}
\label{sect:ics}

\begin{table}
\centering
\begin{tabular}{c c c c}
\hline
Run & $R$($a<$1 au) & $R$(1$<a<$2 au) & $R$(2$<a<$3 au) \\
\hline
NOJS & 3.33 & 1.76 & 0.28 \\	
8JS & 3.81 & 1.56 & 0.22 \\	
GROW & 2.95 & 2.16 & 0.25 \\	
\hline	
\end{tabular}
\caption{Ratios of total embryo mass to total planetesimal masses ($R$) in different disk regions after 3 Myr of evolution for our three different simulations.}
\label{table:R}
\end{table}

Our three runs (NOJS, 8JS and GROW) finish with an average total terrestrial disk mass of 4.3 $M_{\oplus}$.  In general, our simulations predict $\sim$0.3 $M_{\oplus}$ embryos forming at $a<$1.0 au, Mars-massed objects accreting in the proto-Mars region, and several small, Moon-massed embryos growing in the inner asteroid belt region (see figures \ref{fig:mass_comp} and \ref{fig:ic_comp}).  However, we find that the belt region is totally dominated by smaller planetesimals, rather than embryos, at this phase of evolution.  Objects similar in mass to Ceres are quite prevalent throughout the asteroid belt (an average of 571 objects with $M_{Ceres}<M<10M_{Ceres}$), while each system contains only $\sim$5 larger, $\sim$Lunar-massed ($>$0.01 $M_{\oplus}$) embryos.  Table \ref{table:R} summarizes the ratio ($R$) of embryo ($M>$ 0.01 $M_{\oplus}$) to planetesimal mass at different locations within the terrestrial disk at $t=$ 3 Myr.  The disparity between our generated $R$ values throughout the disk and assumptions of N-body studies \citep{chambers01,ray09a,clement18} is an important takeaway from our simulations because different bimodal make-ups can lead to different evolutionary outcomes in different radial regions. We address each zone individually in the following sections.

\subsubsection{Inner Disk}

The innermost section of our disks, where $M_{iso}$ is approached rapidly, are heavily depleted of planetesimals at $t=$ 3 Myr.  This can have significant implications on the follow-on evolution of Earth and Venus analogs within the giant impact phase.  In general, higher $R$ values lead to less dynamical friction generated by planetesimals.  As swarms of planetesimals can damp the orbits of the growing planets (discussed further in section \ref{sect:control}), this can result in final systems of planets with unrealistically large eccentricities and inclinations.  However, \citet{lykawka19} analyzed terrestrial growth within various evolutionary schemes using initial values of $R=$ 1, 4 and 8.  The authors noted that high-$R$ disks were typically more successful at yielding small Mars analogs, replicating late veneer accretion on the Earth, and generating Mercury-Venus pairs.  Furthermore, \citet{jacobson14} found that employing a high-$R$ disk was an effective mechanism for delaying the Moon-forming impact and thus providing an adequate match to the amount of material delivered to form the late veneer. Therefore, a high-$R$ disk would be advantageous if another mechanism were capable of limiting the eccentricities and inclinations of Earth and Venus.
  
  \subsubsection{Mars Region}

In the Mars region, slightly elevated $R$ values \citep[$R\simeq$ 2.0 as opposed to $R=$ 1.0 often assumed in the literature:][]{chambers01,ray09a,clement18} can potentially alter final system outcome within an early Nice Model evolutionary scheme.  \citet{clement18} argued that instabilities timed $\sim$1-10 Myr after nebular gas dispersal are most successful at limiting Mars' mass (compared with earlier instabilities) because the higher $R$ values achieved at more advanced evolutionary stages leads to levels of dynamical friction that are insufficient to save material from loss during the instability. Therefore, in an early instability scenario, the higher values of $R$ in the Mars region (table \ref{table:R}) of our simulations might lead to greater mass loss \citep{clement18_ab} and improved outcomes in a scenario where the giant planet instability is the Mars mass-depletion event.  Moreover, \citet{lykawka19} found that higher-$R$ disks ($R=$ 4 or 8, as opposed to $R=$ 1) were about twice as likely to produce a Mars analog with the correct mass and orbital offset from Earth.

\subsubsection{Asteroid Belt}

Consistent with equation \ref{eqn:tgrow}, the primordial asteroid belt planetesimals are relatively unprocessed at $t=$ 3 Myr.  This is slightly more pronounced in our simulations that include the giant planets (however, we find this trend to be weak and inconclusive; table \ref{table:R}).  While we would expect our results in the inner disk to be somewhat independent of our selection of initial planetesimal mass (as $T_{grow} \lesssim \tau_{gas}$), our asteroid belt results are significantly biased by our initial conditions.  Thus, as accretion events in the asteroid belt are rare in our simulations, our selected primordial asteroid sizes are strongly preserved in the final SFD.  While it should be noted that we do not test this hypothesis by varying $M_{init}$, if the initial asteroids indeed formed large \citep{morby09_ast,johansen15,dermott18}, our results indicate that the region would be dominated by a few Lunar-massed asteroids, and dozens of Ceres-massed objects when the nebular gas dissipated.  However, it should be noted that these results are specific to our assumption of a ``heavy'' primordial belt and a uniform $\Sigma \propto r^{-3/2}$ disk \citep[see, for example,][for an analysis of different disk profiles]{iz14,izidoro15}.

\begin{figure}
	\centering
	\includegraphics[width=.5\textwidth]{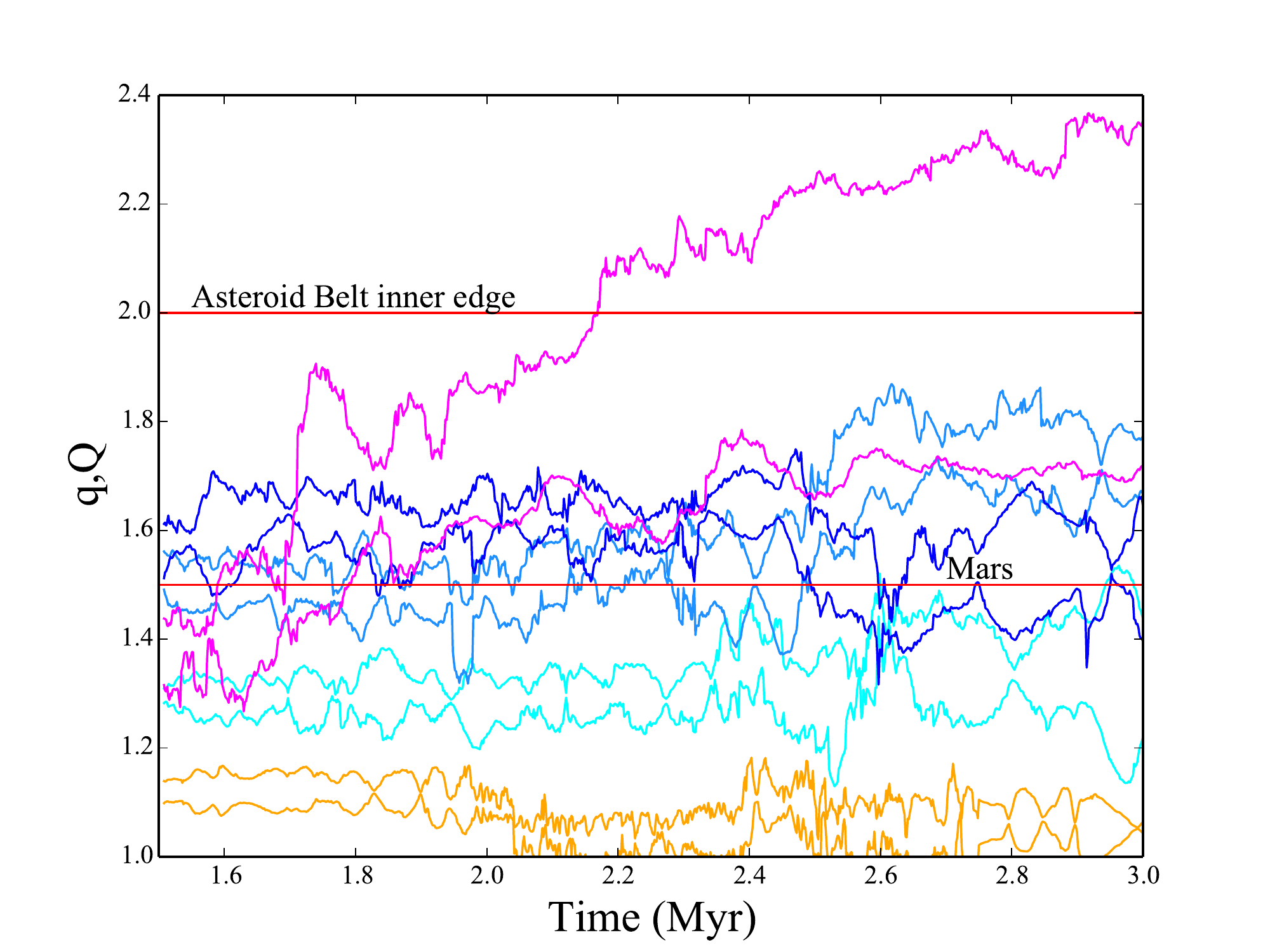}
	\caption{Perihelia and aphelia vs. time for three growing embryos in the Mars region (plotted in different shades of blue; each begins with a masses between that of the Moon and Mercury at $t=$ 1.0 Myr, and grow to around a Mars mass at $t=$ 3.0 Myr) interacting with and scattering a small planetesimal (pink line) into the asteroid belt.  The proto-Earth embryo is plotted in orange.  The horizontal red lines correspond to the modern semi-major axes of Mars and the asteroid belt's inner edge}
	\label{fig:qaq}
\end{figure}

Planetesimals are occasionally implanted into the belt from the inner disk regions in our simulations.  The largest radial migration of a Ceres-massed asteroid from $t=$ 1-3 Myr in any of our simulations is $\sim$1.1 au, and each system implants an average of $\sim10^{-4}M_{\oplus}$ worth of material originating with $a<$ 1.5 au in the belt. Therefore, if Vesta had formed in the inner terrestrial region \citep[as suggested by its composition:][]{bottke06_nat,mastro17}, our results imply that its implantation in the belt could have occurred during the gas disk phase of evolution \citep[however the scarcity of such events in our simulations would indicate that this is unlikely given the subsequent depletion of the belt:][]{obrien07,clement18_ab}.  An example of this type of planetesimal scattering from our GROW simulation is plotted in figure \ref{fig:qaq}.  In that run, a small planetesimal originating at $a=$1.38 au  experiences a series of close encounters with 3 large embryos in the proto-Mars region that drive it's aphelion well in to the asteroid belt.  While the planetesimal's orbit still crosses that of one of three embryos at $t=$ 3 Myr, it is possible that it could be further scattered onto a stable orbit in the asteroid belt during the giant impact phase \citep[e.g.:][]{sandine19}.

\subsubsection{Comparison with Previous Work}

\begin{figure}
	\centering
	\includegraphics[width=.5\textwidth]{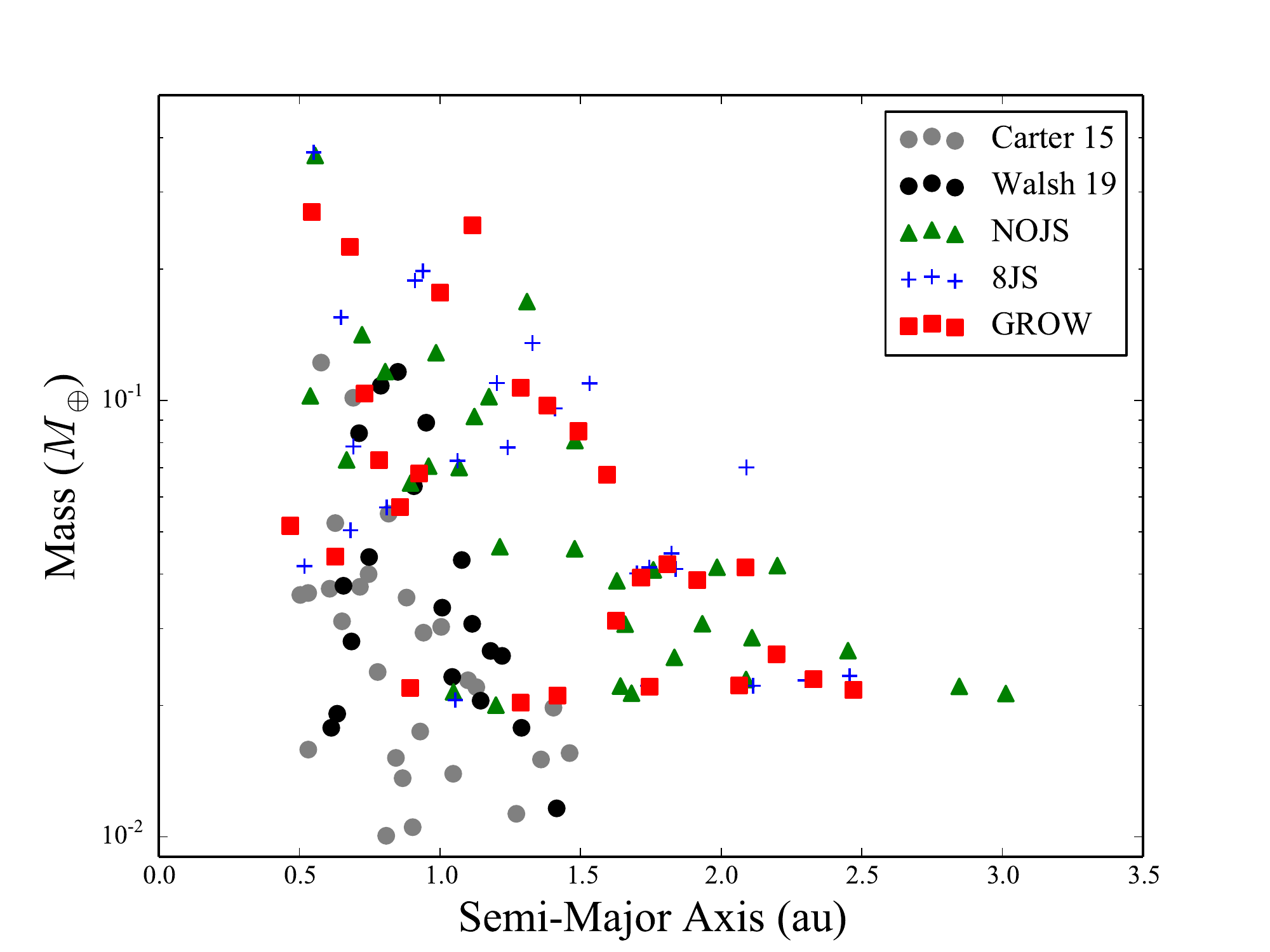}
	\caption{Embryo (only objects with $M>0.01M_{\oplus}$ are plotted) distributions at $t=$ 2 Myr for our 3 different simulations compared with results from \citet{carter15} and \citet{walsh19} (grey and black circles, respectively).}
	\label{fig:mass_comp}
\end{figure}

\begin{figure}
	\centering
	\includegraphics[width=.5\textwidth]{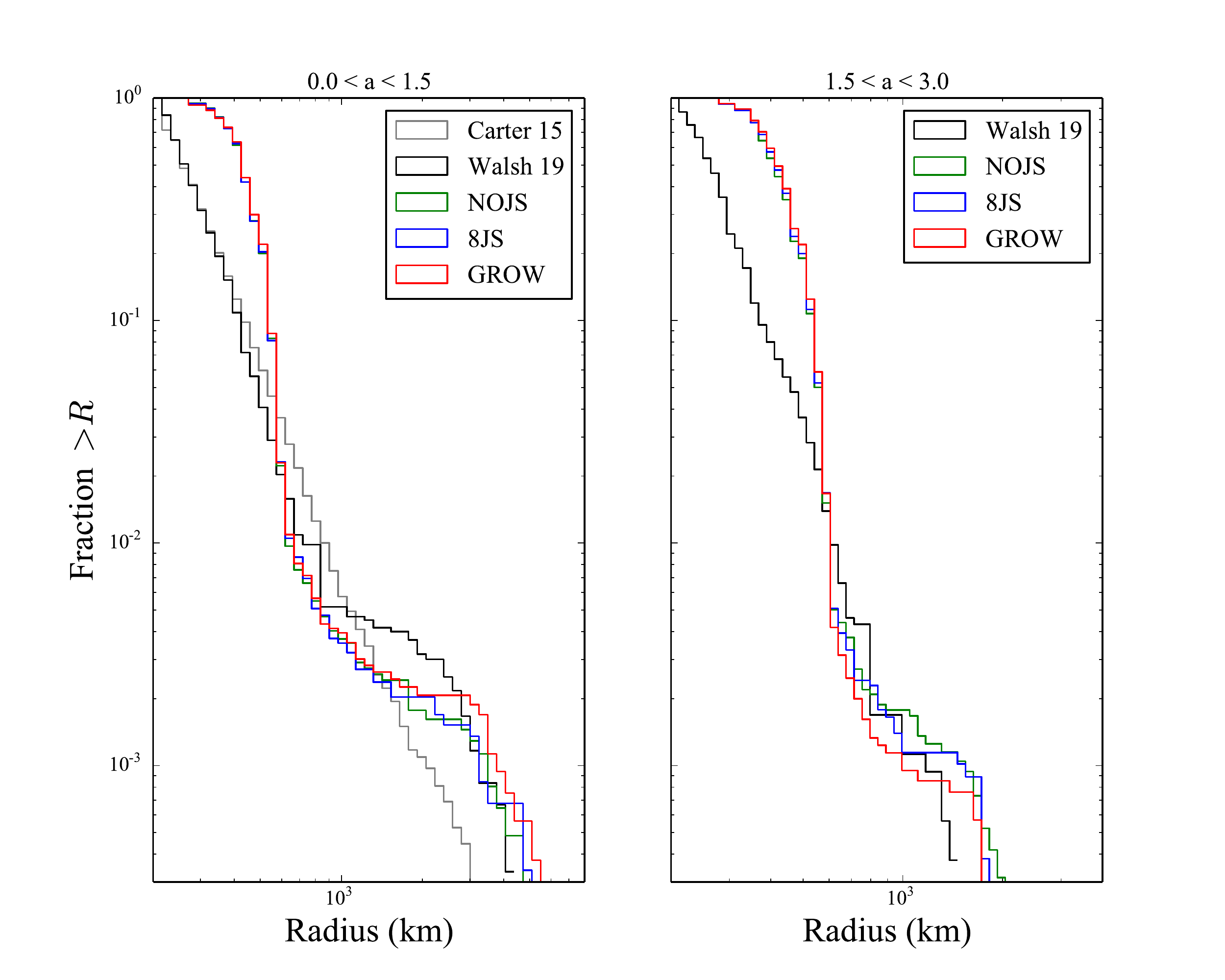}
	\caption{SFD at $t=$ 2 Myr for the inner ($a<$ 1.5 au, left panel) and outer (1.5 $<a<$ 3.0 au, right panel) regions of the terrestrial disk, compared with results from \citet{carter15} and \citet{walsh19} (grey and black lines, respectively).}
	\label{fig:sfd}
\end{figure}

We compare the state of our terrestrial disk at $t=$ 2 Myr with previous work by \citet{carter15} and \citet{walsh19} in figures \ref{fig:mass_comp} and \ref{fig:sfd}.  In both figures, we consider the results of each paper's nominal, MMSN calculations \citep[using the nomenclature of][this is the high-resolution, calm disk]{carter15}.  For reference, the simulation from \citet{carter15} does not include the giant planets, while \citet{walsh19} place 1 $M_{\oplus}$ versions of Jupiter and Saturn at $a=$ 3.5 and 6.0 au that are instantaneously moved to their modern masses and pre-instability orbits \citep[$a=$ 5.0 and 9.2 au:][]{levison11} at $t=$4.0 Myr.  Additionally, the simulation of \citet{carter15} begins with $D=$ 196-1530 km planetesimals drawn from a SFD power-law of $dN = m^{-2.5}dm$ and simplifies collisions by inflating radii by a factor of 6.  In contrast, \citet{walsh19} use initial planetesimals drawn from a distribution centered around $r=$ 30 km.  Because our simulations begin with $D\simeq$ 200 km, equal-mass planetesimals, it is not surprising that our final SFDs (figure \ref{fig:sfd}) are relatively steep, with a tail superimposed by runaway growth.

We note significant differences between our results, and those of both previous studies.  Three major factors contribute to these disparities: our wider initial planetesimal disk, our intra-annulus interpolation method, and our direct treatment of collisions and close encounters without collisional fragmentation.  While \citet{carter15} and \citet{walsh19} study more narrow disk regions (0.5-1.5 and 0.7-3.0 au; respectively), our work considers the entire radial range of 0.48-4.0 au (table \ref{table:new}).  Perhaps the largest difference between our respective final embryo distributions (plotted in figure \ref{fig:mass_comp}) is the prevalence of Moon-Mars massed embryos in the 1.5-2.0 au region, and Moon massed embryos in the inner asteroid belt in our simulations.  These differences are mostly a result of our larger initial planetesimals.  Nominal simulations in \citet{walsh19} begin with $r\simeq$ 15 km planetesimals placed throughout the disk.  Since 2 Myr is significantly less than $T_{grow}$ at $a>$ 2.0 au, our results are more biased by the larger planetesimal sizes than they are by the outcomes of runaway growth.  This is further evidenced by low $R$ values in the asteroid belt in our simulations.  The difference in initial planetesimal sizes is also fossilized in our final SFDs (figure \ref{fig:sfd}), thus resulting in an over abundance of $\sim$100-500 km planetesimals compared with \citet{carter15} and \citet{walsh19}.

In general, our final SFDs significantly more indicative of runaway growth than are those of \citet{carter15} and \citet{walsh19}.  We speculate that this is a result of our treatment of collisions and close-encounters (without inflating planetary radii or utilizing tracer particles).  Thus, once the total planetesimal number decreases, growing embryos in our simulations begin to both gravitationally focus smaller planetesimals onto collision courses, and heat up the local velocity dispersion through scattering events.

Our method of interpolating between annuli likely artificially accelerates growth, and we find this effect to be most consequential in the 0.5 $<a<$ 1.0 region.  When we begin to interpolate at $t=$ 100 Kyr, the oligarch embryo in the annulus 1 ($r=$ 0.5 au) is about five times larger than annulus 2's ($r=$ 1.0 au) oligarch.  As we linearly interpolate between these regions when laying new oligarch embryos (rather than logarithmically), embryos in the middle of the intra-annulus regions are boosted in mass \citep[relative to those of, say,][]{walsh19}.  As this is also true for annuli 2 and 3, this has the cumulative effect boosting the total mass concentrated in embryos in the inner terrestrial disk relative to those of \citet{carter15} and \citet{walsh19}.

\citet{carter15} and \citet{walsh19} each incorporate algorithms designed to account for the effects of collisional fragmentation by introducing new ``fragment'' particles (or tracers) when imperfect collisions occur \citep[e.g.:][]{leinhardt12,stewart12}.  Conversely, our work treats all collisions as perfectly accretionary.  It is difficult to assess the degree to which our results differ from those of models considering collisional fragmentation without a consistent set of control runs.  It is also unclear whether collisional fragmentation plays a significant role in altering the final distribution of embryo masses because generated fragment particles can obviously be re-accreted later in the simulation; thus lengthening the accretion timescale while resulting in a similar final system architecture.  Indeed, \citet{deienno19} concluded that energy dissipation occurring during fragmenting, embryo-embryo collisions does not contribute to significant differences of final system structure in terrestrial planet formation (both in terms of orbital excitation and planet mass).  However, other authors using different numerical implementations have reached the opposite conclusion \citep[eg:][]{bonsor15,clement18_frag,kobayashi19}.  Thus, one could argue that the results of computational investigations of fragmentation are dependent on the specific numerical approach taken (and therefore more sophisticated models are required to study the problem).  While a complete analysis of imperfect accretion is beyond the scope of this work, we cannot discount collisional fragmentation as a potential contributor to the observed differences in embryo masses and planetesimal SFDs in figures \ref{fig:mass_comp} and \ref{fig:sfd}.  Therefore, growth towards larger embryo masses is potentially artificially accelerated in our simulations compared to those of \citet{carter15} and \citet{walsh19} (the degree to which is unclear).

We also note that our disk populations are significantly more bimodal in mass than \citet{carter15}.  At first glance, this seems to imply that our simulations are significantly further evolved within the runaway growth phase.  However, simulations in \citet{carter15} begin with planetesimal sizes ranging from 196-1530 km in order to resemble a more advanced stage of oligarchic growth.  Thus, this range of primordial sizes seems to persist in the simulations of \citet{carter15} through the $t=$ 2 Myr point.  Therefore, the differences between our respective SFD's can be interpreted as a fossilization of the initial planetesimal population.   In section \ref{sect:control}, we speculate further about how this fossilized initial planetesimal distribution might effect the giant impact phase.

\subsubsection{New Distributions for Giant Impact Studies}

\begin{figure}
	\centering
	\includegraphics[width=.5\textwidth]{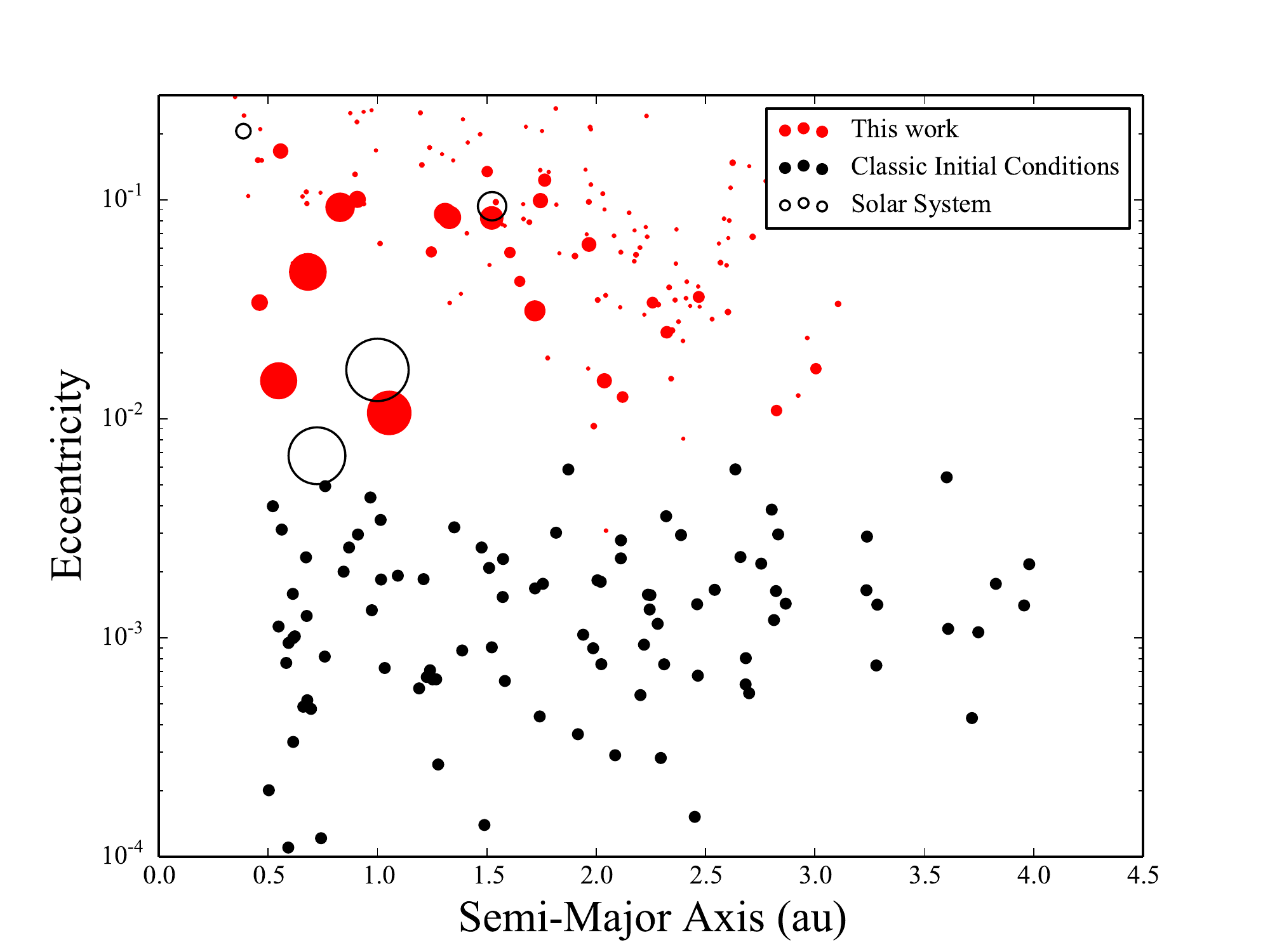}
	\caption{Comparison of all objects with $M>$0.001$M_{\oplus}$ in our GROW simulation (red dots) with a typical distribution of 100 equal massed embryos used in N-body studies \citep[black dots, e.g.:][]{chambers01,obrien06,clement18}, and the modern terrestrial planets (open circles).  The size of each point is proportional to the object's mass.}
	\label{fig:ic_comp}
\end{figure}

In figure \ref{fig:ic_comp} we plot the state of our GROW simulation at $t=$ 3 Myr, compared with typical initial conditions used in simulations of the giant impact phase \citep[e.g.:][]{chambers01}.  In our simulations, the inner disk ($r\lesssim$ 1.5 au) attains quite an advanced evolutionary state prior to nebular gas dispersal, and the system already resembles some aspects of the modern terrestrial architecture.  In fact, the largest two embryos in our 8JS simulation (a proto-Venus analog at $a=$ 0.6 au and a proto-Earth at $a=$ 0.9 au) each possess masses of 0.39 $M_{\oplus}$.  The consequences of such a distribution of embryos for the post-gas disk phase of giant impacts is obvious.  As described in \citet{walsh19} (we expand upon this further in section \ref{sect:tp_form}), the ultimate phase of terrestrial assembly proceeds as a late instability.  Thus, rather than a hundred or so embryos accreting over hundreds of giant impacts, around a dozen embryos experience just a handful of massive impacts as they continue to accrete small bodies over hundreds of Myr.  Furthermore, such an evolutionary scheme might be consistent with planetary differentiation models \citep{rubie15} that suggest that Venus' lack of an internally-generated magnetic dynamo implies that primordial stratification in its core was never disrupted and mixed by a late giant impact \citep{jacobson17b}.

\subsection{Fully Evolved Systems}
\label{sect:tp_form}

\begin{table*}
\centering
\begin{tabular}{c c c c c}
\hline
Set & $N_{TP}$ & $AMD/AMD_{SS}$ & $RMC/RMC_{SS}$ & $M_{Mars}$ \\
\hline
NOJS & 5.1 & 0.73 & 0.29 & 0.63 \\	
8JS & 6.4 & 0.55 & 0.29 & 0.40 \\	
GROW & 5.5 & 1.04 & 0.28 & 0.60 \\	
Classic, 1000 & 4.4 & 5.6 & 0.41 & 0.76 \\
Classic, 2000 & 4.3 & 5.8 & 0.37 & 0.43 \\
\hline	
\end{tabular}
\caption{Summary of various statistics for our different batches of simulations designed to study the giant impact phase of terrestrial planet formation (NOJS, 8JS, GROW, and classic \citep{chambers01} initial conditions with either 1,000 or 2,000 equal-mass planetesimals).  The columns are as follows: (1) the simulation set (each made up of 24 separate integrations), (2) the mean total number of terrestrial planets ($a<$ 2.0 au,  $m>$ 0.05 $M_{\oplus}$) after 200 Myr, (3) the mean normalized AMD (equation \ref{eqn:amd}), (4) the mean normalized RMC (equation \ref{eqn:rmc}) and (5) the mean Mars analog mass \citep[in Earth units, defined here simply as the largest planet with 1.3$<$a$<$2.0 au:][]{clement18}.}
\label{table:tp_form}
\end{table*}

For a first order approximation of our $t=$ 3 Myr systems' evolution up to $t=$ 200 Myr, we perform an additional suite of simplified, CPU integrations where the planetesimal population is approximated with 1,000 equal-mass objects.  We provide a summary of important statistics for each set of runs in table \ref{table:tp_form} \citep[commonly cited as ``success criteria'' for terrestrial planet formation models.  See for example:][]{ray09a,clement18,izidoro18_book_review}.  The evolution of these systems within the giant impact phase is strikingly different from that of the ``classic'' model for terrestrial planet formation \citep[e.g.:][]{wetherill80,chambers01}.  Most notably, our new distributions of large, $\sim$0.1-0.4 $M_{\oplus}$ embryos struggle to combine into systems of four, larger terrestrial planets.  The largest embryos accrete the remaining planetesimals (as well as the occasional smaller embryo), however the embryo systems seldom destabilize fully and experience a final series of giant impacts with one another.  Thus, the resulting systems contain too many terrestrial planets that are systematically under-massed.  Indeed, the mean number of planets with $a<$ 2.0 au and $m>$ 0.05 $M_{\oplus}$ among our 8JS simulations using new initial conditions is 6.4, as opposed to 4.3 in our simulations that employ ``classic'' initial conditions (discussed further in section \ref{sect:control}).  Furthermore, because our new systems evolve only slightly over 200 Myr, the final terrestrial planets largely maintain the dynamically cold orbits that were originally damped via interactions with the gas disk.  To demonstrate this, we calculate the normalized angular momentum deficit \citep[AMD:][]{laskar97} and radial mass concentration statistics \citep[RMC:][]{chambers01} for each system:

\begin{equation}
	AMD = \frac{\sum_{i}M_{i}\sqrt{a_{i}}[1 - \sqrt{(1 - e_{i}^2)}\cos{i_{i}}]} {\sum_{i}M_{i}\sqrt{a_{i}}} 
	\label{eqn:amd}
\end{equation}

\begin{equation}
        RMC = MAX\bigg(\frac{\sum_{i}m_{i}} {\sum_{i}m_{i}[\log_{10}(\frac{a}{a_{i}})]^2}\bigg)
        \label{eqn:rmc}
\end{equation}

\begin{figure}
	\centering
	\includegraphics[width=0.5\textwidth]{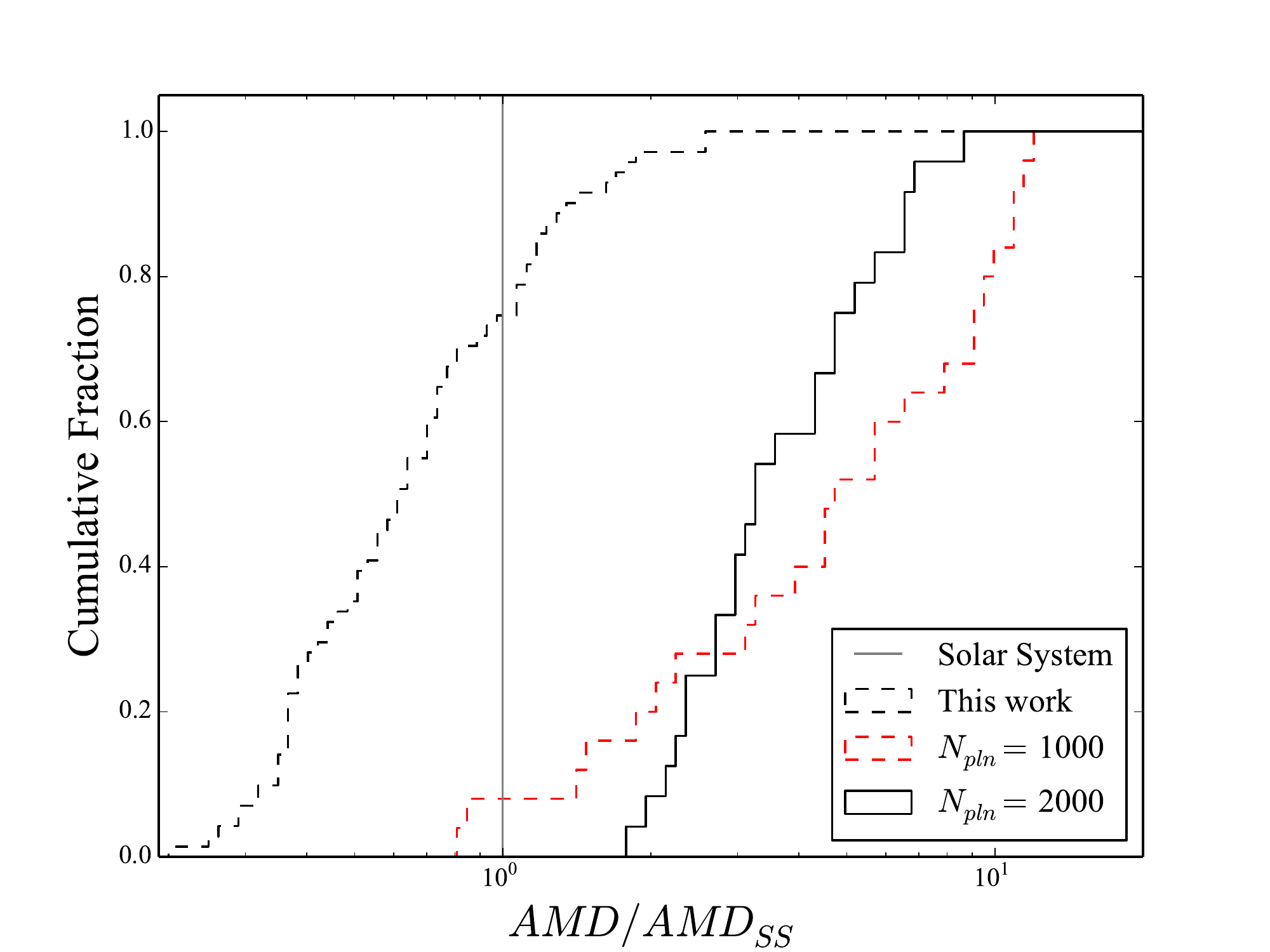}
	\caption{Cumulative distribution of normalized AMDs (equation \ref{eqn:amd}) for three separate batches of terrestrial planet formation simulations.  The black lines plot simulations that employ ``classic'' initial conditions \citep[e.g.:][]{chambers01,obrien06,clement18} where half of the disk mass is placed in 50 equal-massed embryos and either 1,000 (solid line) or 2,000 (dashed line) equal-massed planetesimals.  The red line represents depicts the results of 72 simulations using embryo distributions generated from the GPU simulations presented in this work.  The grey vertical line denotes the solar system AMD for Mercury, Venus, Earth and Mars.}
		\label{fig:amd}
\end{figure}

Figure \ref{fig:amd} plots the cumulative distribution of system AMDs for our 72 simulations that are based off the results of our GPU simulations, compared with 50 control simulations that make use of classic initial conditions.  Given the limited number of large accretion events experienced in our GPU-derived simulations, the final terrestrial architectures consistently provide better matches to actual inner solar system in terms of system $AMD$.  We hesitate to conclude that this result implies a potential solution to the terrestrial over-excitation problem given the poor solar system analogs produced by our integrations.  Specifically, we consistently form under-massed Earth and Venus analogs that are too great in number, and over-massed Mars analogs that are also overabundant.  Nevertheless, the result of final system $AMD$ being limited in systems where embryos attain a more advanced evolutionary state in the gas phase is intriguing, and an avenue for future development and study.

\begin{figure}
	\centering
	\includegraphics[width=.5\textwidth]{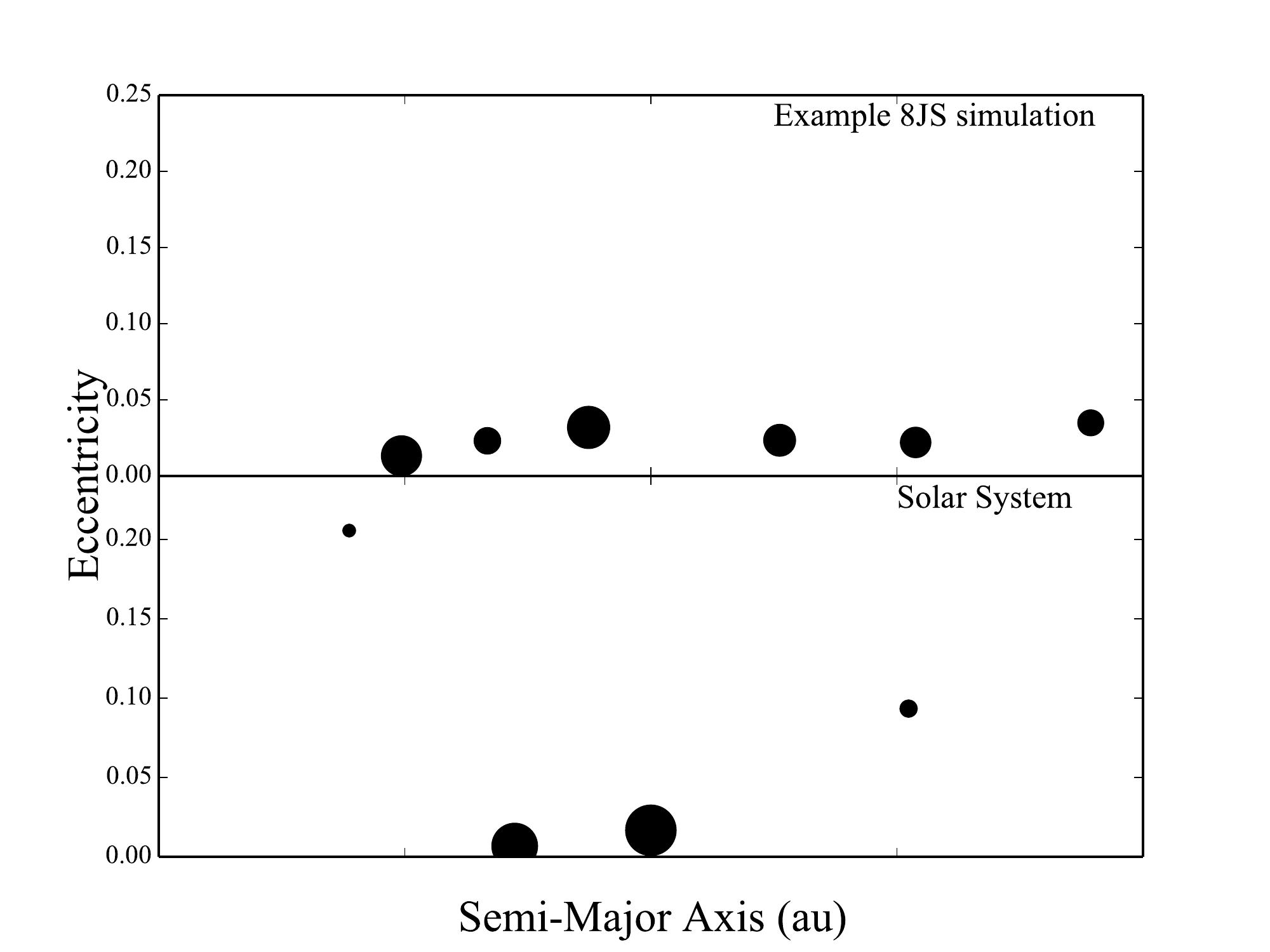}
	\caption{An example final terrestrial system from the 8JS batch (note that this example is chosen specifically to highlight the failure of this model).  In order of increasing semi-major axis, the 6 terrestrial planets in the system have masses of 0.62, 0.26, 0.69, 0.38, 0.35 and 0.25 $M_{\oplus}$, respectively.}
		\label{fig:example}
\end{figure}

At first glance, it would appear that the reason for the stunted growth of our systems of larger embryos is the presence compact MMR chains that develop as a result of aerodynamic drag induced migration during the gas disk phase \citep[a less extreme version of the ``breaking of chains'' model for compact systems of Super-Earths described in][]{izidoro17}.  Indeed, many of the large embryos in our GPU simulations finish near the major first order MMRs.  As an example, the 7 largest embryos with $a<$ 2.0 au in our GROW simulation lie just outside of a mutual 3:2,3:2,5:3,4:3,4:3,4:3 chain.  However, on closer inspection, these proto-planets are not in resonance, nor do they fall into resonance during the giant impact phase.  Instead, the dearth of massive accretion events in these simulations can be attributed to the high $R$ values in the inner disk, in conjunction with a relatively wide spacing between embryos \citep[as opposed to, say,][]{koko_ida00}.  Since the large embryos emerge from the gas disk on low eccentricity orbits well outside one another's mutual Hill spheres, with very little planetesimal mass available to perturb them on to crossing orbits, these systems routinely remain stable for 200 Myr.  An example of such a system from the 8JS set is plotted in figure \ref{fig:example}.  Given the modern eccentricities of Mercury ($e=$ 0.21) and Mars ($e=$ 0.09), the degree of orbital excitation in this system is remarkably low, and a typical outcome of our study.  However, Earth and Venus only attain $\sim$70$\%$ of their modern masses, and four additional planets 2-3 times the mass of Mars are stable in the system (one in the region between Earth and Venus, and three in the Mars region).  

The total mass of planets in the Mars region (as well as that of the largest Mars analog; table \ref{table:tp_form}) is clearly too large in our simulations.  However, this is somewhat expected given that our systems grow from a uniform, 5 $M_{\oplus}$ disk of material.  Thus, we do not account for early depletion in the Mars-forming and asteroid belt regions that might have resulted from giant planet migration \citep{walsh11} or a primordial gap \citep{ray17sci}, nor do we consider the dynamical excitation of the giant planets \citep[which is highly efficient at limiting the mass of Mars, e.g.:][]{ray09a,lykawaka13,bromley17}.  Nevertheless, it is obvious that if the Earth and Venus analogs in figure \ref{fig:example} each accreted one of the additional $\sim$0.2-0.3 $M_{\oplus}$ embryos, perhaps ejecting an additional Mars analog in the process, the final system would provide a much better match to the modern Venus-Earth-Mars architecture.

\subsubsection{Implications for the Moon-forming impact}

It follows naturally to speculate that, given our results, a dynamical trigger might be required to destabilize such a compact system terrestrial embryos, eject additional Mars analogs, form the Moon, and complete terrestrial planet formation.  The logical trigger would be the giant planet instability \citep{Tsi05,levison08,deienno17}.  Several recent authors have invoked an early instability rather than a ``delayed'' instability coincident with the Late Heavy Bombardment \citep{tera74,gomes05}.  \citet{morb18} argued that, since the crystallization of the Moon's mantle took longer than the Earth's, sequestration of highly siderophile elements (HSEs) during the crystalization process can explain the observed Earth-Moon HSE disparity, and is consistent with an early instability.  Indeed, the early impact chronology on the Moon and Mars appears consistent with an instability occurring within 100 Myr of the solar system's birth \citep{mojzsis19,brasser20} Additionally, \citet{nesvorny18} showed that the instability must have occurred within 100 Myr of nebular gas dispersal in order to permit the survival of the Patroclus-Menoetius binary system of Jupiter Trojans.  Furthermore, a delayed dynamical event in the outer solar system is at odds with the recently discovered asteroid families by \citet{delbo17} and \citet{delbo19} in the inner main belt that are inferred to be as old as the solar system \citep[e.g.:][]{milani17}. Finally, an early instability is also capable of limiting Mars' accretion \citep{clement18}, and adequately exciting \citep{deienno18} and depleting \citep{clement18_ab} the asteroid belt.  However, \citet{clement18_frag} requires a very specific timing for the instability (1-5 Myr after gas dissipation) to limit Mars' mass and prevent the terrestrial disk from ``re-spreading'' and forming 3-4 equal-mass planets (however this issue might be less pronounced in a higher-$R$ disk, see discussion in section \ref{sect:ics}).  Imposing such a strict constraint on the instability's occurrence is problematic in that it conflicts with recent studies of Neptune's effect on the Kuiper Belt \citep{nesvorny16} that aim to explain the inclination distribution of the 3:2 MMR population \citep{nesvorny15a}.  Specifically, \citet{nesvorny15b} requires that Neptune migrate smoothly for $\sim$20 Myr before experiencing a ``jump'' in semi-major axis \citep[though recent work by][suggests that other timescales are also viable]{volk19}.  Thus, we propose that, assuming more realistic initial distributions of embryos in the terrestrial forming disk (as produced via our high-resolution GPU simulations) might provide greater flexibility in terms of the timing of the giant planet instability for the early instability scenario proposed in \citet{clement18}.  Indeed, we find that Mars analogs already have masses of order $\sim$0.1 $M_{\oplus}$ after the gas disk phase (figure \ref{fig:emb_compare}).  In several of our simulations of the giant impact phase, many Mars analogs do not grow larger beyond the $\sim$10$\%$ level over 200 Myr (31$\%$ of the planets in the region accrete no additional embryos after $t=$10 Myr).  Thus, it seems reasonable, given our simulation results, that a system of 4-5 Mars-massed planets formed during the gas disk phase \citep[consistent with the hypothesis that Mars is a ``stranded embryo'' given its rapid inferred accretion timescale:][]{Dauphas11,kruijer17_mars} could remain stable for some tens of Myr before disrupted by the giant planet instability.

Our generated embryo configurations also imply a mass ratio between the proto-Earth and the Moon-forming impactor (Theia) closer to unity.  This result is interesting given that a giant impact involving two $\sim$0.5 $M_{\oplus}$ bodies has been shown to be successful at replicating the observed isotope ratios \citep{canup12}.  \citet{kaibcowan15} found these conditions to be highly improbable within dynamical simulations of terrestrial planet formation that invoke classic \citep{chambers01} initial conditions.  While beyond the scope of our present manuscript, the implications of our GPU-evolved embryo populations are nonetheless intriguing with respect to the Moon's formation.

In summary, the results of our simplified simulations that follow the evolution of our GPU-generated embryo distributions within the giant impact phase lead us to speculate that an alternate evolutionary sequence might have ensued during the ultimate phase of terrestrial assembly in the solar system.  However, given the simplicity of the numerical simulations presented in this paper, we leave the full development of this scenario to future work.  In short, our results indicate that the giant impact phase might have played out as a delayed instability, as proposed in a similar study by \citet{walsh19}.  This starkly contrasts the rather prolonged sequence of hundreds of giant impacts that is modeled throughout much of the literature \citep[e.g.:][]{obrien06,fischer14,lykawka19}.  Because the interior regions of our terrestrial disks achieve extremely high $R$ values and possess well-spaced orbital configurations during the nebular gas phase, we propose that a dynamical trigger such as the Nice Model instability is necessary to stimulate the destabilization of the primordial proto-planets in the inner solar system.  In such a scenario, the instability would have to be responsible for both triggering the final few giant impacts on Earth and Venus \citep[most importantly the Moon-forming impact, e.g.:][]{quarles15,kaibcowan15} and evacuating the $\sim$1.3-2.0 au region of additional Mars-massed planets \citep{clement18}.

\subsection{Dependency on Planetesimal Sizes}
\label{sect:control}

\begin{figure}
	\centering
	\includegraphics[width=0.5\textwidth]{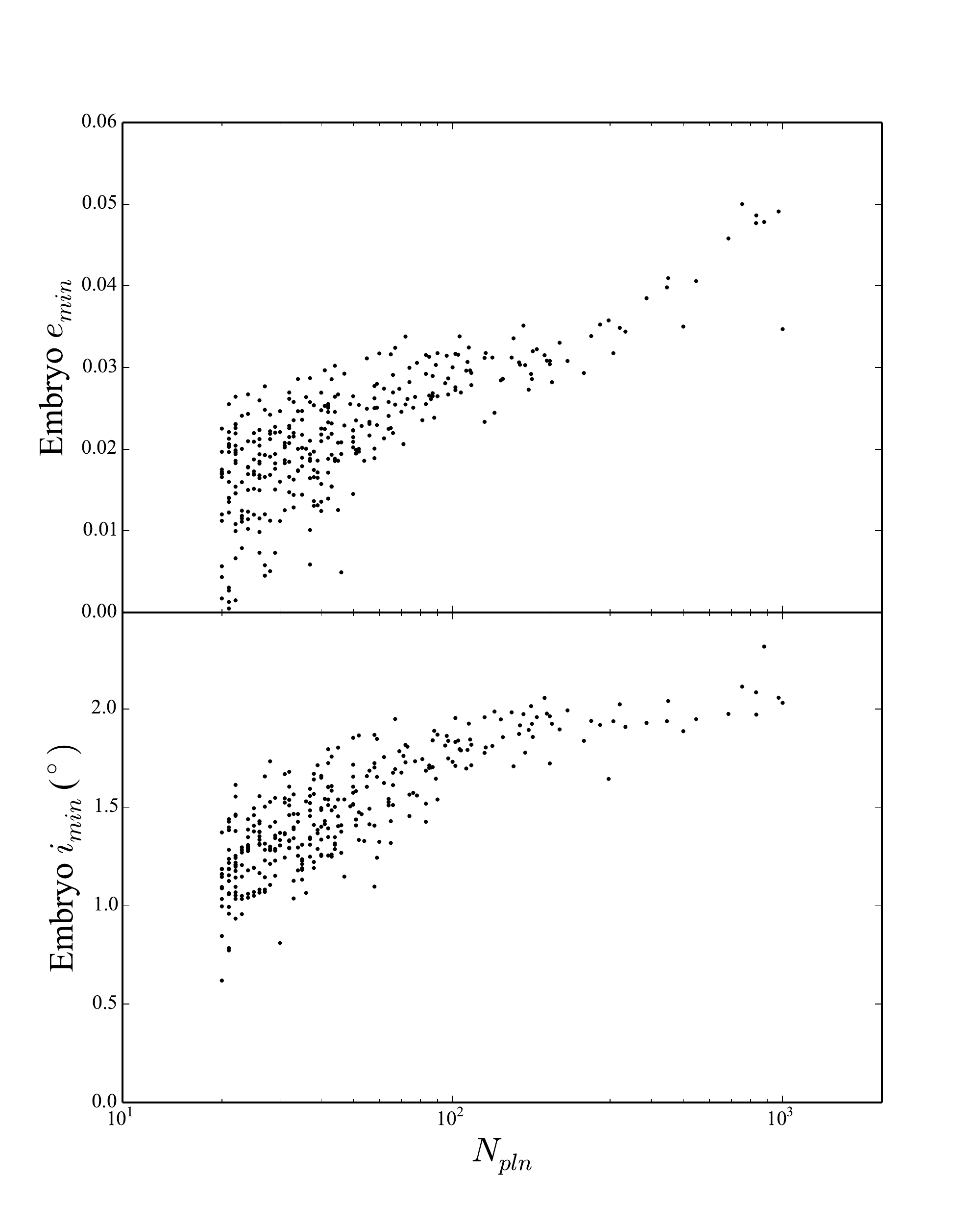}
	\caption{Minimum eccentricity (top panel) and inclination (bottom panel) attained by a 1.0 $M_{\oplus}$ embryo embedded in a disk of $N_{pln}$ equal massed planetesimals for 1.0 Myr.  $N_{pln}$ is varied in each simulation while the total mass of planetesimals is fixed at 1.0 $M_{\oplus}$.}
	\label{fig:test}
\end{figure}

Left behind after the processes of embryo formation and oligarchic growth is a remnant of the initial planetesimal size distribution.  While understanding the properties of the first generation of planetesimals is still an area of active research \citep[e.g.:][]{levison15,draz16,wallace17}, we argue that the SFD of the residual planetesimals can influence the final system AMDs.  As a proto-planet grows within a swarm of smaller planetesimals, it undergoes a constant series of close-encounters that tend to reduce the system's AMD.  Thus, the $e/i$ evolution of a growing planet in a planetesimal disk can be thought of as a random-walk of encounters with a net trend towards damping the planet's orbit.  It follows that, with a smaller number of larger planetesimals, it is possible to randomly walk towards lower vales of $e/i$ and smaller AMDs.  We demonstrate this concept in figure \ref{fig:test} with a simple numerical experiment using $Mercury$ \citep{chambers99}.  In each simulation, we embed a 1 $M_{\oplus}$ planet at 1 au within a distribution of planetesimals with a total mass of 1 $M_{\oplus}$.  We place the large planet on a moderately excited initial orbit ($e=$0.1, $i=$5.0$^{\circ}$), and in all cases the orbit is markedly damped after 1 Myr.  However, simulations using a smaller number of large planetesimals experience significantly greater damping, and display a larger range of outcomes than those with a greater number of small planetesimals \citep[despite the total planetesimal mass remaining fixed.  See also:][]{jacobson14,kobayashi19}.

We continue to test this concept with an additional suite of 50 simulations (see table \ref{table:tp_form}) of the classic terrestrial planet formation model \citep[e.g.:][described in detail in section \ref{sect:nbody}]{chambers01}.  In 25 simulations, the planetesimal population is modeled using 1,000 objects, each with mass $M=$ 0.0025 $M_{\oplus}$.  Our second set of 25 simulations considers 2,000 planetesimals with $M=$ 0.00125 $M_{\oplus}$.  Each batch of simulations finish with nearly identical mean $AMD$ values (0.0101 and 0.0104, respectively, see figure \ref{fig:amd}) that are $\sim$6 times that of the modern solar system.  However, the set of simulations employing fewer, large planetesimals has a greater dispersion of $AMD$ outcomes (min$_{1000}=$ 0.0014, max $_{1000}=$ 0.024, $\sigma_{1000}=$ 0.0072, min$_{2000}=$ 0.0052, max$_{2000}=$ 0.018, $\sigma_{2000}=$ 0.0035, see figure \ref{fig:amd}).  Thus, a terrestrial system forming within a distribution of larger planetesimals is able to randomly walk to both lower and higher AMD values.  If we are to define ``success'' as satisfying a constraint 50$\%$ of the time \citep{nesvorny12},  most terrestrial formation models \citep[with the notable exception of Grand Tack evolutionary schemes;][]{walsh11,walsh16} struggle to consistently replicate the solar system's low AMD.  As we find smaller populations of more massive planetesimals to be more successful at producing low AMD terrestrial planets, we argue that a primordial terrestrial disk of $r\sim$100 km planetesimals (albeit not akin to the planetesimals used in our $N_{pln}=$1,000 simulations) formed directly via gravitational instability \citep[e.g.:][]{morby09_ast,johansen15} is worth investigating.

\section{Conclusions}

We present detailed simulations of embryo formation within a decaying gas disk starting from $r\sim$100 km planetesimals.  Our calculations begin by following oligarchic growth within individual radial annuli.  As the total particle number decreases in each annulus, we interpolate within the intra-annulus regions, and assemble the entire terrestrial disk (0.48$<$a$<$4.0 au) in a single simulation after 1 Myr.  Thus, our results are somewhat biased by our interpolation method (though we find this error term to be minor).  Specifically, future work should employ a logarithmic means of interpolating between different annuli, rather than a linear one.  There are several important takeaways from our work, and that of other recent high-N studies of embryo formation \citep{carter15,walsh19,wallace19}.

\subsection{Bimodal Makeup Depends on Radial Location}

We show that the ratio of total mass in embryos to the total mass in planetesimals ($R$) existing around the time of gas-disk dispersal varies strongly with semi-major axis.  In the Earth/Venus-forming region we find $R$ values as high as $\sim$4.0, as compared with more moderate ratios ($\sim$2.0) in the proto-Mars region, and low  concentrations of embryos ($R\simeq$ 0.20) in the primordial asteroid belt.  We argue that the different values of $R$ in each disk region can lead to a substantial differences in system outcome during follow-on evolution.  For instance, high $R$ values in the inner disk have been shown to increase the probability of forming Venus/Mercury analogs \citep{lykawka19}, while more moderate ratios in the Mars-forming region can potentially help limit Mars' final mass in an early Nice Model scenario \citep{clement18}.

\subsection{Few Giant Impacts in the Giant Impact phase}

Perhaps the most striking difference between our generated distributions of embryos and planetesimals \citep[as well as those from similar studies:][]{carter15,walsh19} and those supposed in classic terrestrial formation models \citep{chambers01,ray09a} is the advanced evolutionary state attained in the $a\lesssim$1.0 au region during the nebular gas phase.  In our simulations, only a handful of reasonably large (0.1 $\lesssim M \lesssim$ 0.4 $M_{\oplus}$) embryos grow in the Earth and Venus forming regions of the disk.  Therefore, the giant impact phase of evolution ensues as a delayed instability \citep[e.g.:][]{walsh19}, with Earth and Venus experiencing only a few giant impacts en route to attaining their modern masses.  Given the limited growth experienced by such embryos in an additional, simplified suite of simulations of the giant impact phase, we speculate that a dynamical trigger \citep[the Nice Model instability:][]{Tsi05,nesvorny12,clement18} is required spur on the ultimate series of impacts in the inner solar system.

\subsection{Large Primordial Planetesimals Generate More Extreme AMDs}

Finally, we perform an additional suite of test simulations to demonstrate the effects of the fossilized primordial planetesimal SFD on final terrestrial angular momentum deficit (AMD).  Planetesimal-embryo encounters tend to damp the orbits of proto-planets via a random walk towards lower eccentricities, inclinations, and total system AMDs.  We show that larger encounters generated from a distribution of fewer, more massive planetesimals allow a system to randomly walk towards both higher, and lower values of AMD.  Thus, we speculate that a primordial generation of massive planetesimals \citep[$r\sim$100 km, formed via gravitational instability][]{morby09_ast,johansen15} might be advantageous in the ultimate giant impact phase of terrestrial assembly in terms of more consistently yielding systems with solar-system like final AMDs.

Our GPU simulations required nearly two years to complete on $NVIDIA$ GK110 (K20X) "Kepler" accelerators, and represent close to the highest contemporaneous resolution achievable with a direct N-body algorithm.   We have shown that the primordial sizes of planetesimals are somewhat fossilized at the end of the gas disk phase; therefore implying that the selection of a particular initial particle mass can lead to significant differences in final system outcomes.  Thus, it is imperative that future authors continue to push the limits of particle resolution as advances in computing power make such endeavors feasible.

\section*{Acknowledgments}

M.S.C. and N.A.K. thank the National Science Foundation for support under award AST-1615975.  NAK also acknowledge support under NSF CAREER award 1846388.  This research is part of the Blue Waters sustained-petascale computing project, which is supported by the National Science Foundation (awards OCI-0725070 and ACI-1238993) and the state of Illinois. Blue Waters is a joint effort of the University of Illinois at Urbana-Champaign and its National Center for Supercomputing Applications \citep{bw1,bw2}.  Further computing for this project was performed at the OU Supercomputing Center for Education and Research (OSCER) at the University of Oklahoma (OU).  This work used the Extreme Science and Engineering Discovery Environment (XSEDE), which is supported by National Science Foundation grant number ACI-1548562. Specifically, it used the Bridges system, which is supported by NSF award number ACI-1445606, at the Pittsburgh Supercomputing Center \citep[PSC:][]{xsede}. Additional computation for the work described in this paper was supported by Carnegie Science's Scientific Computing Committee for High-Performance Computing (hpc.carnegiescience.edu).

\bibliographystyle{aasjournal}
\newcommand{\sci}{$Science$ }
\bibliography{embryo.bib}
\end{document}